\documentclass[12pt,preprint]{aastex}

\usepackage{mathptm}
\usepackage{graphicx}
\usepackage{url}
\usepackage{color}
\usepackage{textcomp}
\usepackage{pdflscape}
\usepackage{mathrsfs}
\usepackage{bm}

\begin{document}

\title{Three-dimensional forward-fit modeling of the hard X-ray and the microwave emissions of the 2015-06-22  M6.5 flare}

\author{Natsuha Kuroda$^1$, Dale E. Gary$^1$, Haimin Wang$^{1,2}$, Gregory D. Fleishman$^1$, Gelu M. Nita$^1$, Ju Jing$^{1,2}$}

\affil{1. Center for Solar-Terrestrial Research, New Jersey Institute of Technology, University Heights, Newark, NJ 07102-1982, USA}

\affil{2. Big Bear Solar Observatory,  40386 North Shore Lane, Big Bear City, CA 92314}

\email{nk257@njit.edu}

\begin{abstract}
The well-established notion of a ``common population'' of the accelerated electrons simultaneously producing the hard X-ray (HXR) and the microwave (MW) emission during the flare impulsive phase has been challenged by some studies reporting the discrepancies between the HXR-inferred and the MW-inferred electron energy spectra. The traditional methods of their spectral inversion have some problems that can be mainly attributed to the unrealistic and the oversimplified treatment of the flare emission. To properly address this problem, we use a Non-linear Force Free Field (NLFFF) model extrapolated from an observed photospheric magnetogram as input to the three-dimensional, multi-wavelength modeling platform \textit{GX Simulator}, and create a unified electron population model that can simultaneously reproduce the observed HXR and MW observations. We model the end of the impulsive phase of the 2015-06-22 M6.5 flare, and constrain the modeled electron spatial and energy parameters using observations made by the highest-resolving instruments currently available in two wavelengths, the Reuven Ramaty High Energy Solar Spectroscopic Imager (RHESSI) for HXR and the Expanded Owens Valley Solar Array (EOVSA) for MW. Our results suggest that the HXR-emitting electron population model fits the standard flare model with a broken power-law spectrum ($\rm{E_{break}}\sim$200 keV) that simultaneously produces the HXR footpoint emission and the MW high frequency emission. The model also includes an ``HXR invisible'' population of nonthermal electrons that are trapped in a large volume of magnetic field above the HXR-emitting loops, which is observable by its gyrosynchrotron (GS) radiation emitting mainly in MW low frequency range.

\end{abstract}

\keywords{Sun: flares -- Sun: HXR, microwave -- Sun: high-energy electrons -- Sun: simulation}

\section{Introduction}

It has been recognized that microwave (MW) emission and hard X-ray (HXR) emission observed during impulsive phases of solar flares show very similar temporal behaviors \citep[][and references therein]{2011SSRv..159..225W}. This signature in general suggests that these two emissions, although observed in very different spectral windows, are produced by a common population of particles under a process of energization during the impulsive phase of solar flares. The dominant emission mechanism in the two wavelengths is likely to be different; the HXR emission is produced by bremsstrahlung and the MW emission is mostly produced by gyrosynchrotron (GS) radiation. The former is produced when high-speed electrons lose energy by collisions with more-stationary targets within the ambient plasma, producing photons with HXR energies. Thus the HXR emission is mostly dependent on and tells us about the ambient plasma density at flare site and the energy of accelerated electrons that collide with them. The latter is produced when moving electrons gyrate due to the Lorentz force in the magnetized plasma. Thus the MW emission is mostly dependent on the energy of mildly relativistic electrons, and the strength of magnetic fields around which these electrons gyrate; i.e., the magnetic field strength at the flare site.

Considering the two emission mechanisms, one may assume that the two observations should converge to the same energy spectrum for the accelerated electrons, and in some simple (single-loop) flares, this is the case  \citep[e.g.,][]{2016ApJ...816...62F}. However, there is a record of observations that seems to suggest otherwise; the indices of nonthermal electron energy spectra inferred from HXR and MW observations are different \citep{1994ApJS...90..599K,2000ApJ...545.1116S}. Generally, the studies that found a difference between the HXR-inferred electron energy spectral index and the MW-inferred electron energy spectral index found that the latter is harder than the former by $\sim$2 \citep{1994ApJS...90..599K,2000ApJ...545.1116S}. Some suggest that considering two different electron energy distributions residing in different energy range (i.e., there is a break in the spectrum) could solve this problem, since MW-emitting electrons are thought to have higher energy than those emitting in HXR \citep{1972SoPh...26..151T,1994ApJS...90..599K}. There are some difficulties with inferring the electron energy spectrum from the observed HXR and MW spectrum, however. On the observational side, flares that enable us to invert the observed HXR photon spectrum to the electron energy spectrum extending above a few hundreds of keV have to be relatively large (e.g. high M- or X-class flares) to obtain enough photon counts above the background level at high energy, and large flares complicate the HXR spectral inversion because the thermal part of the HXR spectrum dominates the nonthermal part at lower energies, sometimes up to 30 keV or higher, thus leading to the inaccurate calculation of the number of nonthermal electrons. There have been some observations of giant flares showing the HXR photon spectrum extending to several hundreds of keV, but the results are mixed regarding the possible break in electron energy spectrum from event to event; the 1980 June 4 event introduced in \citet{1988SoPh..118...49D} showed a hardening break of $\sim$2  at $\sim$300 keV in the HXR photon spectrum spanning from $\sim$20 keV to $\sim$20 MeV, while 2002 July 23 event observed by the Reuven Ramaty High Energy Solar Spectroscopic Imager (RHESSI) up to 8.5 MeV did not show such a break \citep{2003ApJ...595L.111W,2003ApJ...595L..81S,2003ApJ...595L..69L}. A ``cold'' 2002 March 10 flare shows a break-down energy spectrum with the break energy $\sim$100 keV in the photon spectrum \citep{2016ApJ...822...71F}. For MW, the frequency resolution of most of the instruments that record the total solar radio intensity spectrum has been limited, and it has been sometimes difficult to determine an accurate turnover frequency, resulting in uncertainties in the spectral index of the optically thin part of the MW spectrum that is used to determine the nonthermal electron energy spectral index \citep{1994ApJS...90..599K,2003ApJ...595L.111W,2009SoPh..260..135K}. The difference in the source locations of HXR and MW emissions (footpoint and/or above-the-loop-top for the former, whole-loop and/or footpoint for the latter) further complicates the simultaneous analysis. On the modeling side, for MW, inverting the electron energy spectrum from the observed MW spectrum has in the past been oversimplified for quantitative analysis. Therefore, the shape of the energy spectrum of the “common” population of electrons producing HXR and MW could be different from event to event, and whether or not it has a break at some energy remains inconclusive.

In this study, we create one unified multi-loop electron population model \footnote{The model and the documentation explaining how to reproduce the results presented in this paper are available at: \url{http://www.ioffe.ru/LEA/SF_AR/models/3dmodels.html}} that can simultaneously reproduce the observed HXR and MW images and spectra at one point in time during a flare. Unlike many previous studies, we use a realistic three-dimensional magnetic field data cube based on magnetic field measurements, positioned at the actual location of the active region at the time of the observation. We constrain the model by using observations from the highest-resolving instruments available in both HXR and MW wavelengths. In Section 2, we introduce these instruments and the HXR and the MW observations obtained from them. In Section 3, we introduce the modeling platform and the workflow of our simulation for the flare. In Section 4, we present the results of the simulation, with quantitative parameters of the electrons. In Section 5, we discuss what the results suggest in terms of the spatial and the energy distributions of high energy electrons in the flare. We note that, since this simulation is a forward-fitting simulation, the model presented in this study is one of many possible solutions, although we endeavor to create the simplest possible model, with the minimum number of electron populations needed to fit the data, as will be shown in Section 3.

\section{Data}

\subsection{Instruments}

In this study, we use four observational data sources to constrain the HXR and MW emission model: HXR images, HXR spatially integrated spectrum, MW interferometric observations, and MW spatially integrated spectrum. We also employ an observation-based three-dimensional magnetic field model. For HXR images and spatially integrated spectrum, the data from RHESSI are used. RHESSI is capable of producing HXR images and spectra at photon energies from 3 keV to 400 keV with angular resolutions as fine as 2 arcseconds, a spectral resolution of $\sim$1 keV and a temporal cadence of typically 4 s (the spacecraft rotation period, although shorter times are possible using demodulation techniques). For this study, the collimators 2 and 4 were excluded due to their insensitivity to below $\sim$20 keV and the lack of segmentation, respectively, and the sensitivities of all the other collimators were lowered to as low as 76 \% of the launch value (the lowest being the collimator 1). The MW data are taken from the newly expanded solar-dedicated radio array the Expanded Owens Valley Solar Array (EOVSA). Formerly known as OVSA (Owens Valley Solar Array), EOVSA is currently being commissioned to have the unprecedented imaging spectroscopic capability in frequency range of 2.5-18 GHz at more than 300 frequency channels,  with the spatial resolution of $\sim$60 arcsec/$\nu_{GHz}$ (finest $\sim$3.3 arcsec at 18 GHz), 1-s time cadence, and four polarizations. At the time of the event used for this study, EOVSA was recording total radio flux intensity and cross-correlated amplitudes from 9 baselines from 7 antennas. We used the total intensity MW spectrum (spatially integrated spectrum at 162 frequency channels within 2.5-18 GHz), and the relative visibilities \citep{1989ApJ...339.1115G} calculated from the cross-correlated amplitudes as a spatial constraint replacement for images, which were not available yet.

The cross-correlated amplitudes from an interferometric baseline can be converted into one-dimensional relative visibility spatial information to determine the characteristic source size in the direction of the baseline orientation, assuming a simple Gaussian source shape. We use the following relationship (see Appendix A for derivation):

\begin{equation}
\ln(V_{Rel}) = \ln\left(\frac{a_{ij}}{\sqrt{{a_{ii}a_{jj}}}}\right) = -8.393 \times 10^{-11} B_{\lambda}^2 d^2 = -9.325 \times 10^{-14} B_{cm}^2 d^2 f_{GHz}^2
\end{equation}

\noindent
where $a_{ij}$ is the cross-correlation amplitudes from the baseline consisting of antenna $i$ and $j$, $a_{ii}$ and $a_{jj}$ are the auto-correlation total power amplitudes from antenna $i$ and $j$, respectively, $B_{\lambda}$ is the projected baseline length measured in wavelengths, $d$ is the one-dimensional characteristic source size in arcseconds, $B_{cm}$ is the projected baseline length in cm, and $f_{GHz}$ is the observing frequency in GHz. Note that, with this definition, $V_{Rel}$ is independent of calibration because antenna-based gains cancel within $\frac{a_{ij}}{\sqrt{{a_{ii}a_{jj}}}}$. This shows that if the plot of $\ln\left(\frac{a_{ij}}{\sqrt{{a_{ii}a_{jj}}}}\right)$ vs. $B_\lambda^2$ shows a linear dependence with a negative slope, then the one-dimensional characteristic source size can be estimated by a simple relationship. The steeper the slope, the larger the source size in the direction of that baseline's orientation. In reality, the source could be elongated more in one direction than another. Such size variance will be projected onto the baselines with different orientation angle; the relative visibility plots from different baselines with different orientations can reveal the characteristic source size in different directions. At the time of the event for this study, the prototype correlator was not producing valid auto-correlations. Therefore, we used the cross-correlated amplitudes from the shortest baseline as a proxy for auto-correlated total power amplitudes for each baseline (i.e., $\sqrt{a_{ii}a_{jj}} \sim a_{12}$ where antenna 1 and antenna 2 constituted the shortest baseline within the array). The length of this baseline at the time of the observation was $B_{\lambda}\sim$508 at the highest frequency (18 GHz), which yields the minimum fringe spacing of $\sim$400 arcseconds. This is confirmed to be well above the size of the target active region, which means that this baseline should not be resolving any flaring sources. Therefore, the cross-correlated amplitude from this baseline can be used as a good approximation for the total power from each antenna, although this approximation now has the disadvantage that $V_{Rel}$ is no longer independent of calibration. We use the frequency dependence of baseline length for individual antenna pairs to convert $V_{Rel}(\lambda)$ to $V_{Rel}(B_\lambda)$, and will compare the observed and simulated visibilities at different frequencies.  This approach accounts for the possible variation of source size with frequency.

For the magnetic field model, we used the Non-Linear Force Free Field (NLFFF) model extrapolated from the SHARP Active Region Patches’ Cylindrical Equal Area (CEA) photospheric vector magnetogram from Helioseismic and Magnetic Imager \citep[HMI;][]{2012SoPh..275..229S} on board the Solar Dynamics Observatory \citep[SDO;][]{2012SoPh..275....3P} (available every 12 minutes). The net force and torque in the observed photospheric field are first minimized by a preprocessing procedure in order to obtain the chromosphere-like data that meets the force-free condition \citep{2006SoPh..233..215W}; for advantages and disadvantages of this approach, see \citet{2017ApJ...839...30F}. The “weighted optimization” method \citep{2000ApJ...540.1150W, 2004SoPh..219...87W} is then applied to the preprocessed photospheric boundary to perform the NLFFF extrapolation within a box of $256 \times 180 \times 200$ uniform grid points, corresponding to $\sim230 \times 160 \times 180$ Mm$^{3}$. The performance of the extrapolation was verified by visually comparing the model field lines with the 171$\AA$ channel images from Atmospheric Imaging Assembly \citep[AIA;][]{2012SoPh..275...17L} on board the SDO.

\subsection{Observations}

\begin{figure}
\figurenum{1}
\epsscale{1.0}
\plotone{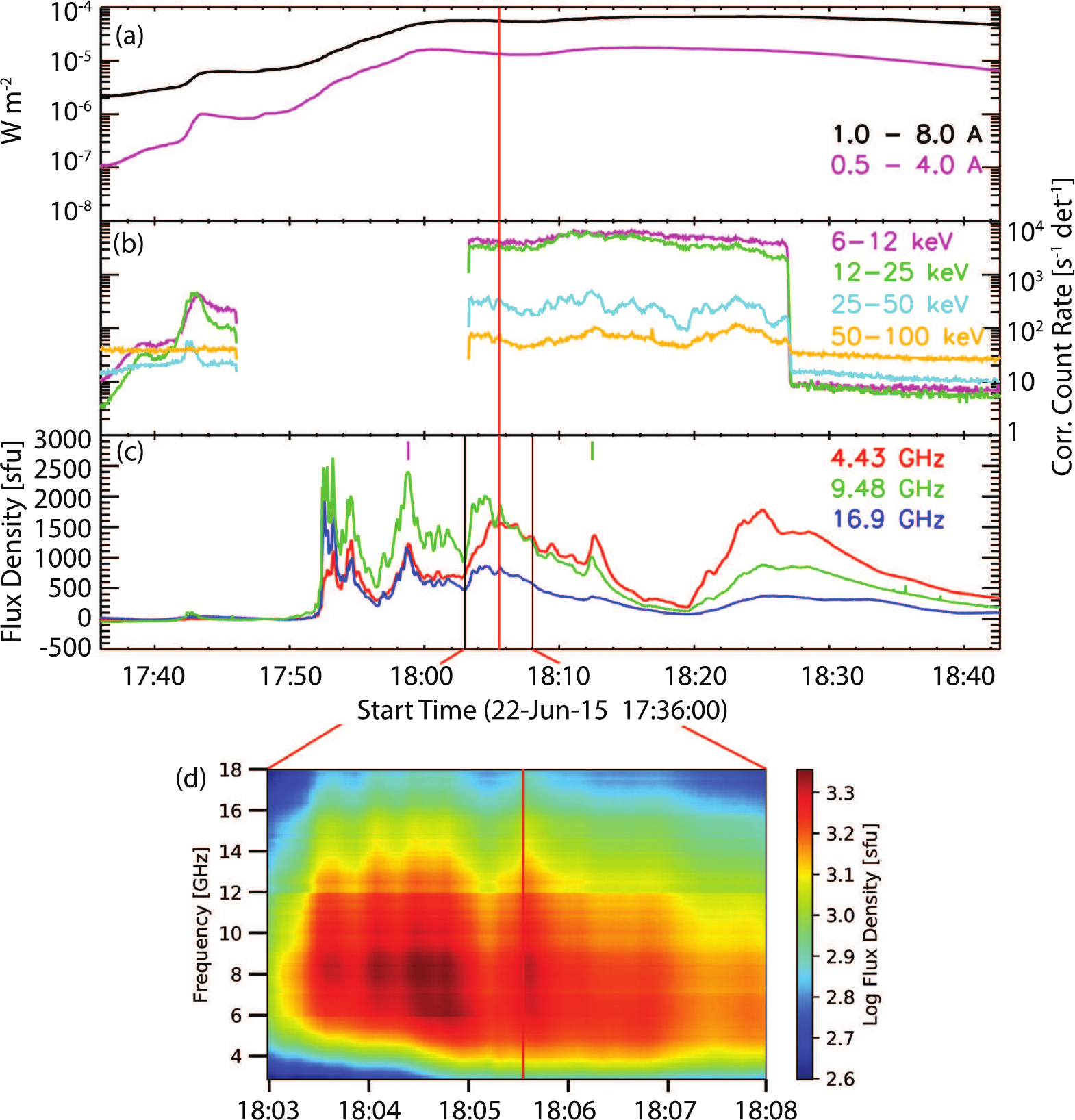}
\caption{(a,b,c) Lightcurves from GOES (a), RHESSI (b), and EOVSA (c) for 2015-06-22 X6.5 flare. RHESSI's data gap is due to its passage through the SAA, and the drop at $\sim$18:27 UT is due to the spacecraft's night time. The vertical red line indicates 18:05:32 UT, the time at which the simultaneous modeling of HXR and MW observations was conducted. The total intensity spectra taken at this time, at 17:58:48 UT (short vertical magenta line), and at 18:12:28 UT (short vertical green line), are shown in Figure \ref{f3}(c). (d) A portion of the EOVSA dynamic spectrum (162 frequency channels in the 2.5 - 18 GHz range, 1 second time resolution) corresponding to the 18:03 UT - 18:08 UT time range indicated by the two vertical black lines in the lightcurve plot above.  \label{f1}}
\end{figure}

In this study, we analyze a well-observed M6.5 flare that occurred on 2015 June 22 \citep{2016NatSR...624319J,2016NatCo...713104L,2017NatAs...1E..85W}. The flare started around 17:50 UT, and reached the Soft X-ray (SXR) peak (1.0-8.0 $\AA$ channel of Geostationary Operational Environment Satellite (GOES) X-ray monitor) at around 18:23 UT, and subsided to background level at around 02:00 UT on the next day. Figure \ref{f1} (a,b,c) shows the lightcurves from GOES, RHESSI, and EOVSA for this flare from 17:36 UT to 18:43 UT. There were two precursors at around 17:24 UT and 17:42 UT \citep[only the latter is shown in Figure \ref{f1}]{2017NatAs...1E..85W}. The parent active region 12371 was located at N13W06 at the time of the flare.

As seen in Figure \ref{f1}(b), RHESSI missed most of the impulsive phase of the flare due to its passage through the South Atlantic Anomaly (SAA). However, EOVSA coverage during this period shows multiple sharp peaks leading to GOES SXR maximum. In this study, we must choose a time covered by both RHESSI and EOVSA for the model, and since HXR images are crucial in this modeling, as will be shown in a later section, we chose the earliest time at which RHESSI's HXR 25-50 keV lightcurve showed the peaky behavior after the spacecraft came out of SAA. This was 18:05:32 UT, which is indicated by the red vertical line in Figure \ref{f1}. The total power dynamic spectrum from EOVSA containing this peak, from 18:03 UT to 18:08 UT, is also shown in Figure \ref{f1}(d). Its smooth time and frequency variation suggest that the emission is dominated by GS mechanism over the entire frequency spectrum.  The dominance of GS emission is expected at these frequencies based on statistics of microwave bursts  \citep{2004SpWea...211005N, 2004ApJ...605..528N}, which show that coherent emission mechanisms are rare above 2.6 GHz.

Figure \ref{f2}(a) shows RHESSI image contours (50, 70, 90\%) in 6-12, 20-35, and 50-75 keV obtained at 18:05:32 UT, integrated over 2 minutes, reconstructed with the CLEAN algorithm, overplotted onto the HMI line-of-sight (LOS) magnetogram taken at 18:04:26 UT. The image shows clear double footpoint emissions in 50-75 keV rooted at regions of opposite magnetic field polarity. The region joining these footpoints is filled by a 6-12 keV source, presumably coming from the loop filled with heated chromospheric plasma, although it cannot be determined if this loop is footed exactly at the 50-75 keV sources. The intermediate-energy 20-35 keV image shows an interesting morphology; two of the sources are nearly co-spatial with the 50-75 keV footpoint sources, but one is located between them, giving the appearance of three sources tracing out one loop that connects them. The centroid of this “middle” 20-35 keV source and the centroid of 6-12 keV source (the center of its northeastern bulge) seems to be slightly shifted, and since there seems to be no region of strong magnetic field corresponding to the location of the “middle” 20-35 keV source (in contrast to the case of 50-75 keV source), we interpret this source as the so-called above-the-loop-top (ALT) HXR source \citep[e.g.,][]{1994Natur.371..495M}. We estimated the difference between the centroid of this source and the centroid of the 6-12 keV source based on the field line geometry of the NLFFF extrapolation model, and found that it is about 20,000 km. This is in agreement with past studies that measured the size of the current sheet formed between coronal HXR source and thermal looptop source in the flare of similar magnitude \citep[M1.2, 17,500 - 33,000 km;][]{2003ApJ...596L.251S}. It is possible that this source is a thermal source, and we will briefly discuss the variation in the model corresponding to this interpretation in later section (Section 4.3). Figure \ref{f2}(b) shows the background-subtracted (background time range was 17:11 - 17:18 UT) RHESSI HXR photon spectrum created by Object Spectral Executive\citep[OSPEX;][]{2002SoPh..210..165S} software, integrated over 8 seconds centered at 18:05:32 UT. The spectral fit was done using OSPEX, with a single-temperature thermal bremsstrahlung radiation function (``\textit{vth}'') and the thick-target nonthermal bremsstrahlung with an isotropic pitch-angle distribution (``\textit{thick2}''). The fitted parameters and goodness-of-fit value are listed in Table 1.

\begin{figure}
\figurenum{3}
\epsscale{.45}
\plotone{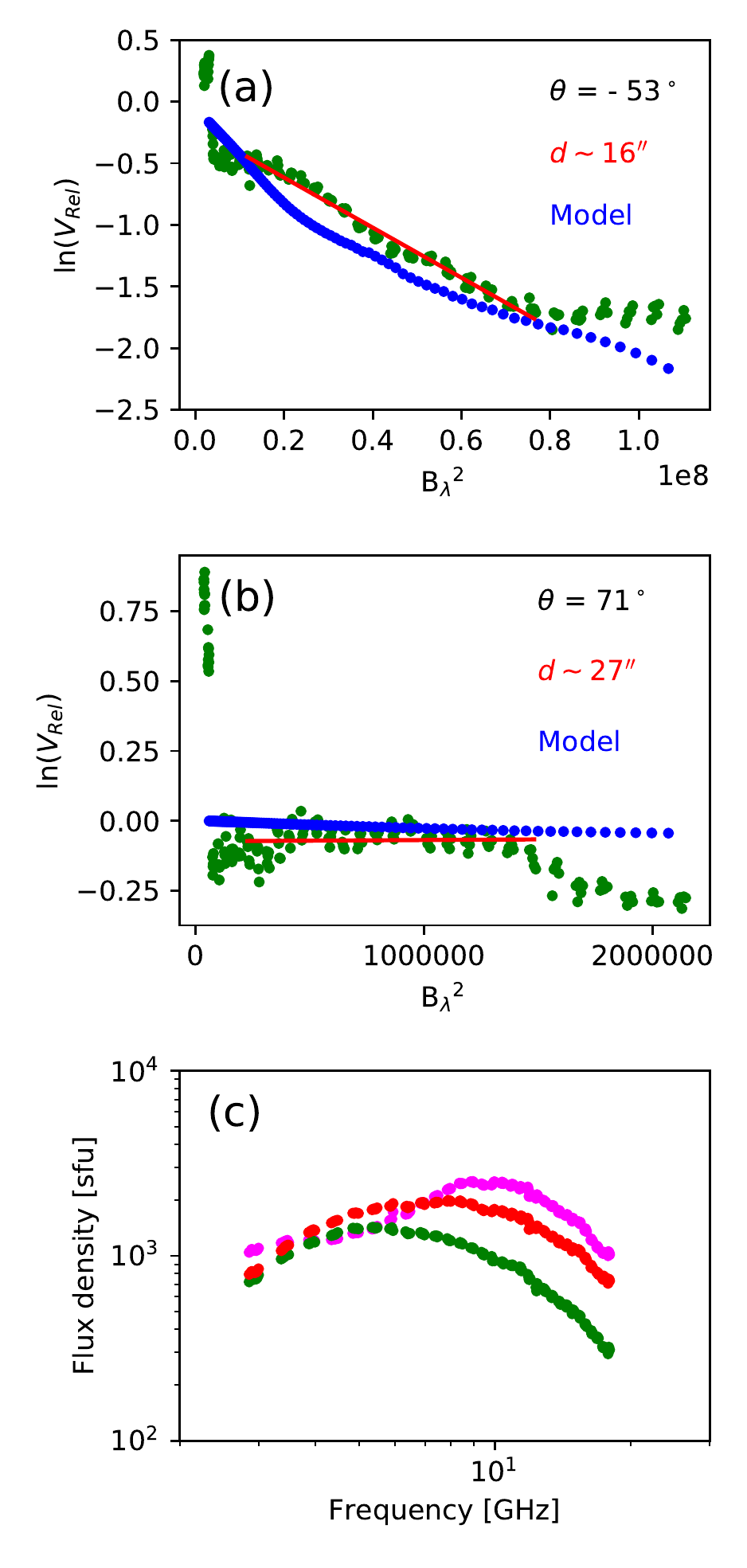}
\caption{(a,b) EOVSA's $\ln(V_{Rel})$ vs. $B_\lambda^2$ plot calculated from the cross-correlated amplitudes taken from the longest baseline (a) and one of the shorter baselines (b) that were available on 2015, Jun. 22. The straight negative slope can be used to calculate the characteristic source size in the direction of  the baseline orientation ($\theta$, clockwise from the Heliocentric-Cartesian x-axis). Note that y values above zero are not considered in the analysis (see Section 2.2). The red lines are the least-squared fits to the ranges of $B_{\lambda}^2$ determined by eye (each corresponding to $6 < f_{GHz} < 15$ and $8 < f_{GHz} < 15$, respectively, as $B_{\lambda}$ has a one-to-one correspondence with frequency). $d$ is the one-dimensional characteristic source size calculated from those fitted slopes. Blue curves are the ones calculated from the model (described later). (c) EOVSA background-subtracted total intensity spectrum plot taken at 17:58:48 UT (magenta), 18:05:32 UT (red, modeled time), and 18:12:28 (green). These times are also marked by three vertical lines (magenta, red, and green) in Figure \ref{f1}(c). \label{f3}}
\end{figure}

Figure \ref{f3}(c) shows the background-subtracted (background's time range was 17:47:43 UT - 17:48:33 UT) MW total intensity spectrum taken from EOVSA at 18:05:32 UT (red). The spectra before and after this time, 17:58:48 UT and 18:12:28 UT, indicated by short magenta and green vertical lines Figure \ref{f1}(c), are also shown in magenta and green, respectively.  Figure \ref{f3}(a) green curve shows $\ln(V_{Rel})$ vs. $B_{\lambda}^2$ plot from the longest baseline, which is able to resolve the smallest feature among all baselines, and \ref{f3}(b) green curve shows one of the shorter baselines with different orientation angle. As introduced in the previous section, if this plot shows a linear dependence, then from the slope we can estimate the one-dimensional characteristic source size in the direction determined by the baseline orientation; -53 degrees for the former and 71 degrees for the latter, measured clockwise from the Heliocentric-Cartesian x-axis. Note that y values above zero (at the low $B_{\lambda}^2$ end) are not considered in the analysis for both baselines, since the numerator of $\frac{a_{ij}}{\sqrt{{a_{ii}a_{jj}}}}$ should not be larger than the denominator in general. We believe that they are coming from the effect of dissimilar calibration at those frequencies among four antennas involved. The longer baseline shows a linear dependence in $10^7 <  B_{\lambda}^2 <  7 \times 10^7$ (corresponding to $6 < f_{GHz} < 15$, as $B_{\lambda}$ has a one-to-one correspondence with frequency as shown in Eqn. 1), whose slope indicates a size of $\sim$16 arcsec (the least-square fit to the range determined by eye). There seems to be a steeper slope in lower $B_{\lambda}^2$ range, although it is difficult to determine the exact source size because of the scatter in the data points (estimates vary between $\sim$20 to $\sim$35 arcsec depending on the fit). The plot above $B_{\lambda}^2 \sim 7 \times 10^7$ looks rather flat or even increasing, which would indicate that there is a new source with competing intensity but different characteristic size appearing above this frequency. Since such judgment is not possible unless we extend the plot beyond the high frequency limit of the instrument, we will not use this part of the relative visibility plot for the observational constraint for our model. The shorter baseline also shows a relatively straight slope in $4 \times 10^5 <  B_{\lambda}^2 <  1.5 \times 10^6$ ($8 < f_{GHz} < 15$), which corresponds to the source size of $\sim$27 arcsec. In summary, these two plots tell us that our target source has a simple, slightly elongated shape from $\sim$8 to $\sim$15 GHz, and perhaps slightly larger size in $\lesssim$ 6 GHz in -53 degree direction.

\section{Simulation Platform and Workflow}

As mentioned in the Introduction, in this study we create a sophisticated 3D model that places within a realistic three-dimensional magnetic field data cube a set of electron populations that reproduce observed HXR and MW emissions. The simulation platform we use is the \textit{GX Simulator} \citep{2015ApJ...799..236N, 2017ApJ-submitted}. It is an IDL-based, graphical-user-interface (GUI) platform that has a highly diverse functionality of which we employ here the following.

First, we import an externally-defined NLFFF magnetic field model into the simulator. Then, we investigate the magnetic field topology and create magnetic flux tubes using the observed HXR images for guidance. This flux tube construction process is an iterative process, and we try to find the magnetic field model that contains the field lines that best connect the observed HXR 50-75 keV footpoint sources by eye. After several trial-and-error iterations, we selected the model cube extrapolated at 18:24 UT. The models at other times, for example, the one at 18:00 UT, had too much shear in the overall field line geometry, and we could not obtain the desired source connectivity. Since 18:24 UT is later than our modeling time (18:05:32 UT), this cube must contain more post-reconnection loops, which should be more suitable for modeling MW emissions that presumably come from trapped electron populations. The flux tubes are developed using these central guiding field lines.

\begin{figure}
\centering
\figurenum{2}
\epsscale{1.0}
\plottwo{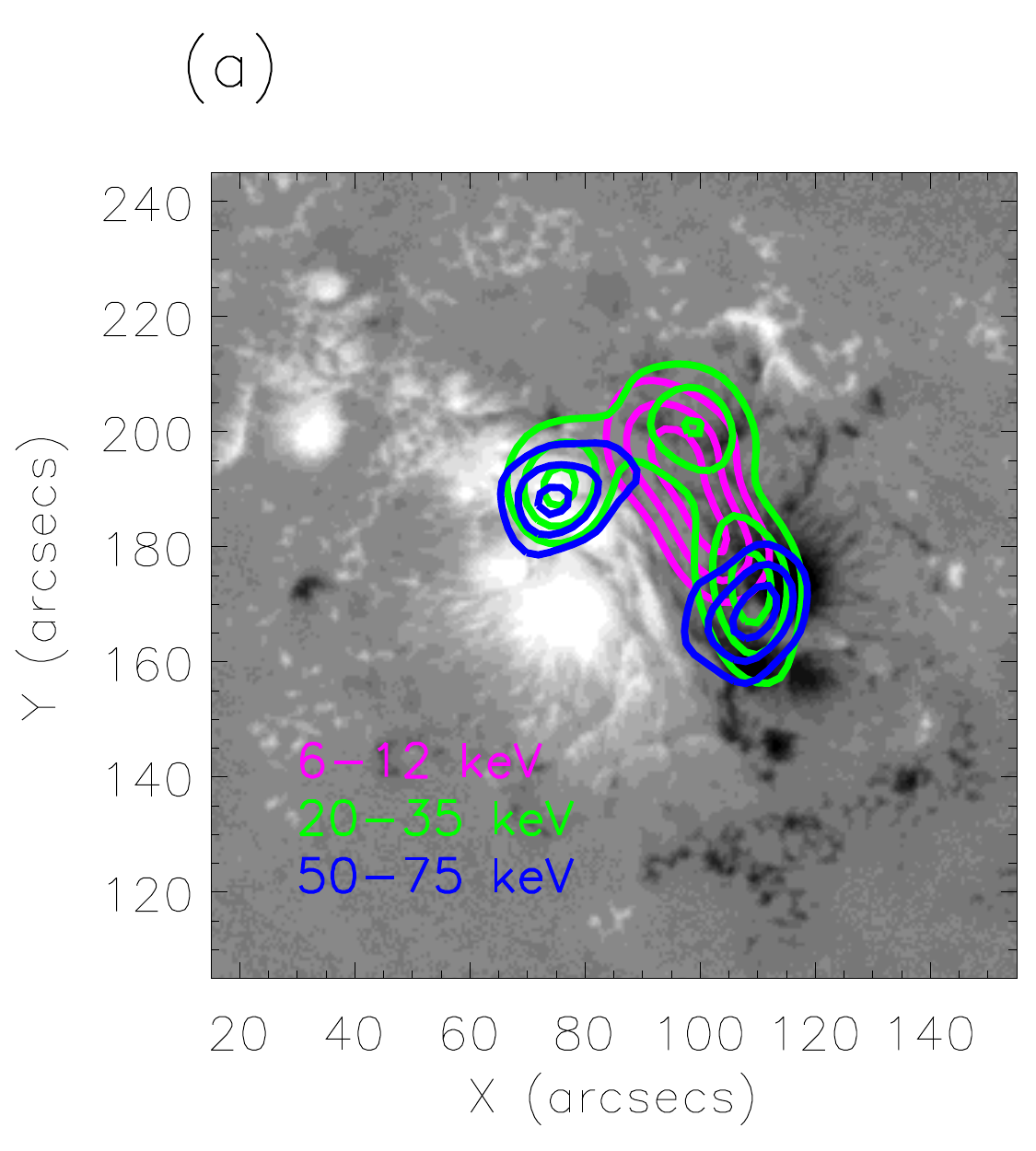}{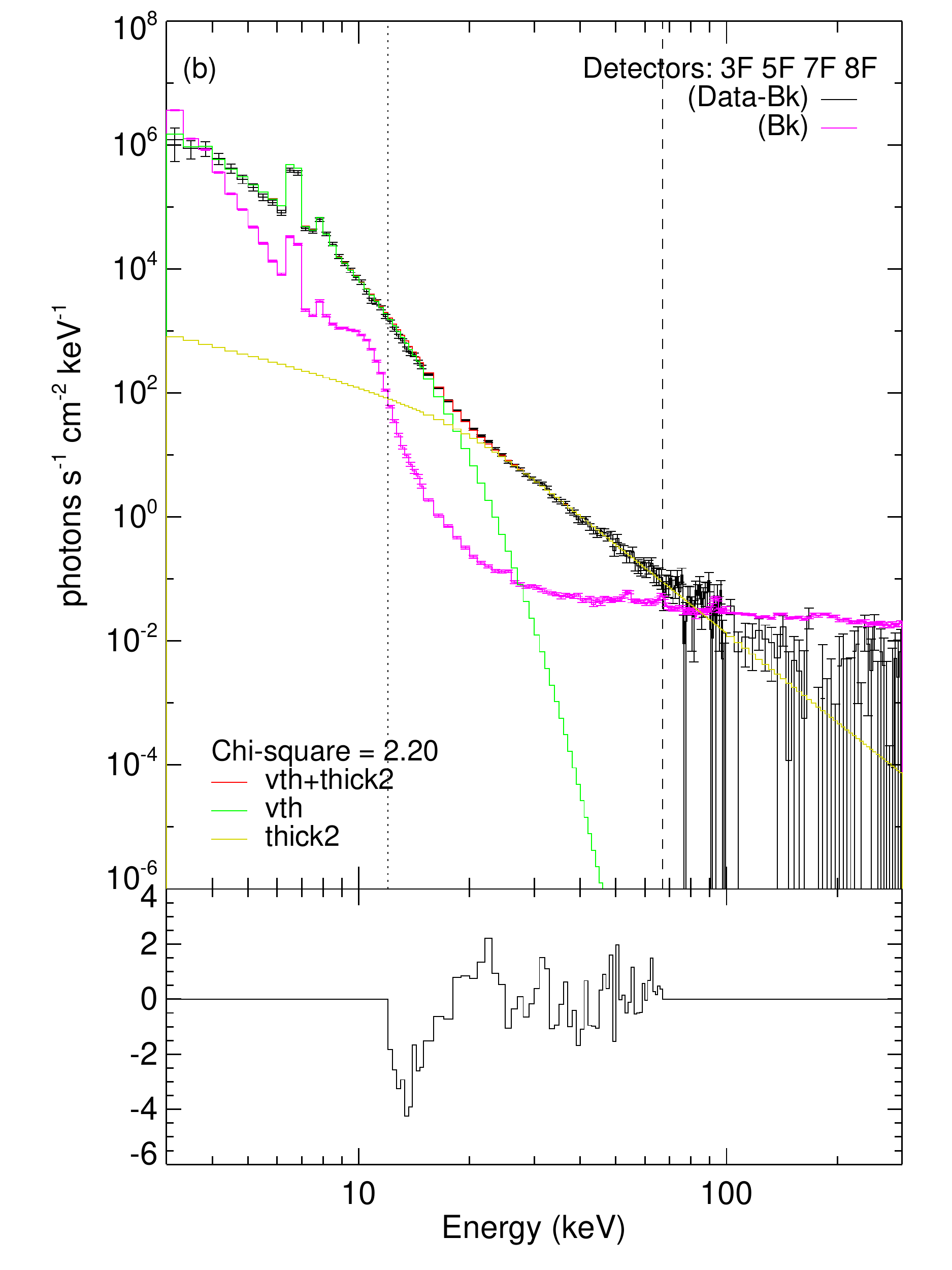}
\caption{(a) RHESSI CLEAN image contours (integrated over 2 minutes) at 18:05:32 UT, overplotted onto the HMI LOS magnetogram taken at 18:04:26 UT. (b) RHESSI HXR photon spectrum taken at 18:05:32 UT, accumulated over 8 seconds using the front-end of collimators 3, 5, 7, and 8. The background was taken from 17:11 UT to 17:18 UT. The spectral fit was done combining \textit{vth} and \textit{thick2} functions in 12-67 keV (dashed lines). The goodness-of-fit value was 2.20, with the normalized residuals plotted at the bottom. \label{f2}}
\end{figure}

Next, we populate the flux tube with thermal and nonthermal populations of electrons, defined to have the required spatial and energy distribution functions. Here, the spatial distributions of the electrons are assigned based on the observed HXR images in Figure \ref{f2}(a), and the energy distribution functions are assigned guided by the OSPEX's spectral fit to the observed HXR spectrum in Figure \ref{f2}(b). For spatial distribution, we use the default functions provided by the simulator (see Appendix B). In cases where some modification is necessary, the simulator allows users to define the electron spatial distribution functions separately for thermal and nonthermal electron populations. For particle energy distribution, the simulator provides a predefined list of well-known functions such as thermal-plus-single-power-law, thermal-plus-double-power-law, Kappa, and others, as well as the choices of isotropic or anisotropic pitch-angle distribution \citep[full list provided in][]{2010ApJ...721.1127F,2015ApJ...799..236N}. The simulator also has an ability to define chromospheric layers with parameters such as plasma density, temperature, and depth. For this study, the initial parameters for the energy distribution for each of the electron populations were taken from the RHESSI spectral fit that was introduced in Section 2.2, which is a thermal-plus-double-power-law consisting of single-temperature thermal bremsstrahlung and the thick-target nonthermal bremsstrahlung radiation functions. The thermal and nonthermal parts were appropriately assigned based on our interpretation of the nature of the observed sources.

Then, we generate 2D HXR and MW images and spectra through calculations using internal codes. The simulator's HXR code calculates the observable flux of HXR photons at 1AU by summing the combination of thermal and nonthermal bremsstrahlung radiations from each voxel. The thermal part of the X-ray code calculates the total bremsstrahlung power radiated from the plasma with a single temperature $T$, taking into account collisions with hydrogen and other atoms, free-free and free-bound transitions, and various line emissions \citep{2002SoPh..210..165S}. The nonthermal code uses the instantaneous bremsstrahlung expression for voxels located at coronal height, and the thick-target bremsstrahlung expression for voxels located at the height of transition region. The expression for the former is:

\begin{equation}
I(\epsilon, \vec{r}) = \frac{n_{p}(\vec{r})V}{4\pi R^2} \int_{\epsilon}^{\infty} Q(E,\epsilon) F(E, \vec{r}) dE
\end{equation}

\noindent
where $n_{p}$ is the plasma proton density, $R$ is 1AU distance, $V$ is the volume of the voxel, $\epsilon$ is photon energy, $F(E, \vec{r})$ is the electron flux density distribution over energy in the given voxel, and $Q(E,\epsilon)$ is the angle-averaged bremsstrahlung cross-section introduced by \citet{1997A&A...326..417H}. The expression for the latter adopts the OSPEX representation of the thick-target model, which is based on the theory described in \citet{1971SoPh...18..489B}, but without the consideration of anisotropic pitch-angle distribution. The electron number density in each interface voxel is converted to the electron flux $\mathscr{F}(E)$ via a simple relationship $\mathscr{F}(E) = F(E,\vec{r})Av(E)$, where $A$ is the surface area of the top of the highest chromospheric voxel and $v(E)$ is the velocity of the electron with energy $E$. Then, provided that the HXR emission is optically thin, the individual contributions are added up along the LOS to form a set of HXR images at various energies. The emission can then be integrated over the image to yield the total power HXR spectrum for comparison with Figure \ref{f2}(b). Currently, the HXR code only calculates electron-ion bremsstrahlung and does not account for Compton scattering or photoelectric absorption of HXRs in the solar atmosphere, with the latter known to produce a broad hump on the photon spectrum around 30-50 keV \citep{1978ApJ...219..705B}. 


The simulator's radio emission code is based on the fast GS algorithm developed by \citet{2010ApJ...721.1127F}, which accounts for GS and free-free radio emission and absorption in a thermal plasma (e.g. Razin effect is included) within the modeled cube (vacuum outside). The fast GS code is a generalization of a numerical Petrosian-Klein (PK) approximation of the exact GS equations, which is more precise than that approximation and also valid for an anisotropic pitch-angle distribution. It can reproduce discrete harmonic structures at low frequencies if requested by the user, or averages over them otherwise. The 2D image at a given frequency is calculated by solving the radiative transfer equation along the LOS, and it includes frequency-dependent mode coupling in addition to emissivity and absorption.

\begin{figure}
\figurenum{4}
\epsscale{1}
\plotone{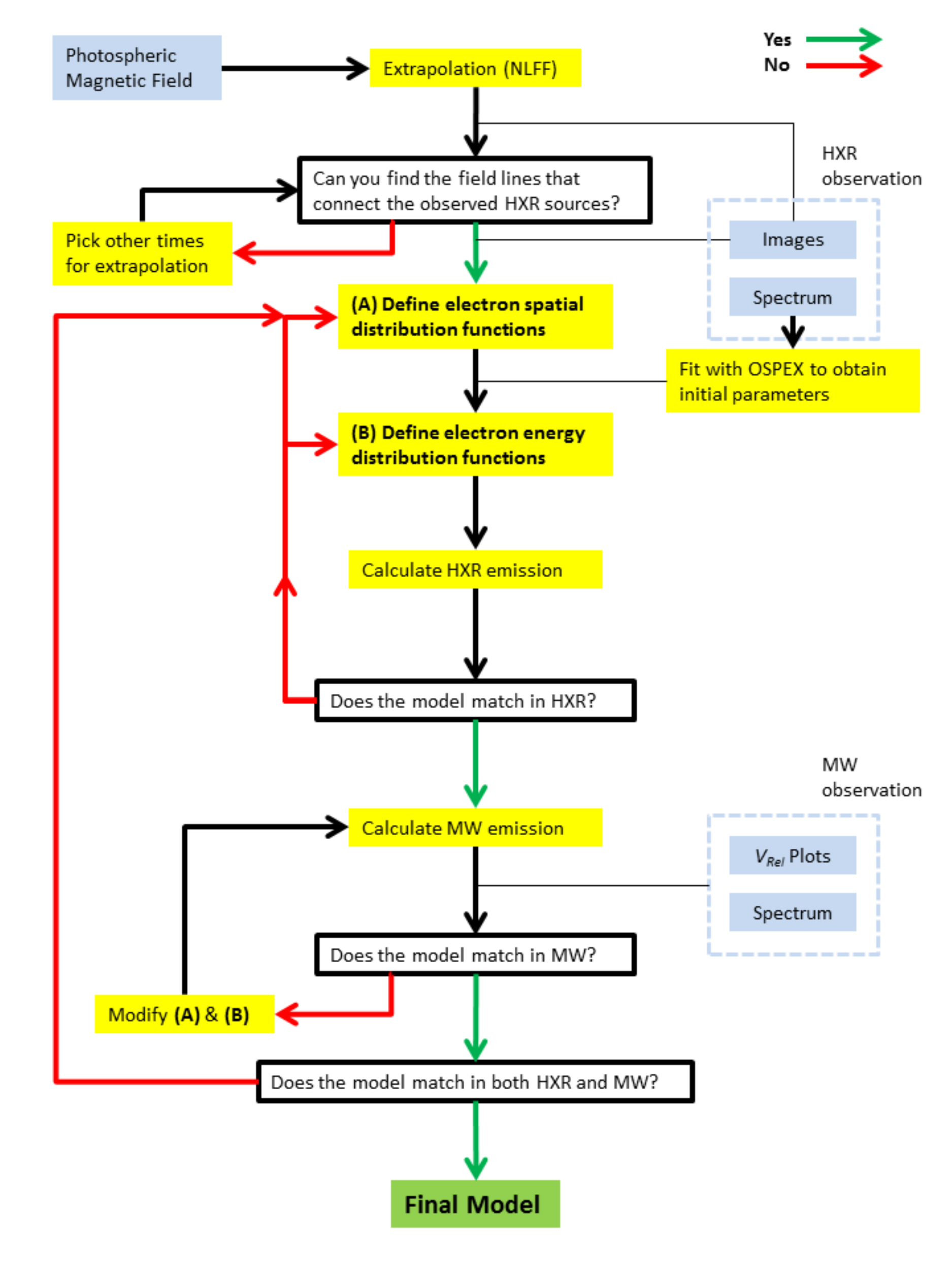}
\caption{The diagram illustating the workflow of the simulation in this study, based on the framework introduced by \citet{2013SoPh..288..549G}. \label{f4}}
\end{figure}

Lastly, we compare simulated HXR and MW images and spectra with the observed images and spectra. The simulator has an ability to convolve the pixelated 2D model image with a user-defined point-spread-function/beam, which enables the user to directly compare the model images to the observed images that go through instrumental responses. For HXR, the model images are convolved with a Gaussian point-spread-function with FWHM of 6.79 arcseconds, according to the nominal FWHM resolution of the finest resolving collimator that was used to reconstruct the observed RHESSI image (collimator 3). For MW, the model visibility was obtained by convolving the visibility of the pixelated model image from the simulator with the sampling function of the EOVSA array at the modeling time. We fine-tune the model in HXR first, and then in MW. The HXR model images and spectra are first created and tested against the observed HXR emission spectra and images by visual inspection. The parameters that are allowed to vary during the fine-tuning process are plasma temperature and spatial distribution parameters for the thermal population, and spatial and energy distribution parameters for the nonthermal population (see Table 2). During the fine-tuning process, the emission measure is constrained by comparing the emission measure calculated by the simulator (the square of thermal particle density integrated over the model volume) with the emission measure calculated from the OSPEX ``\textit{vth}'' function. After fine-tuning this HXR-constrained model, we use it to produce the model MW images and spectra, which are later tested against the observed MW relative visibility and spatially-integrated spectrum. The same parameters are allowed to vary, but in this step, we have the freedom to choose an additional flux tube if necessary, as long as the relative visibility calculated from the model images agree with the observed relative visibility. The modeled relative visibility is compared to the observed relative visibility by eye. The modeled spectrum is fine-tuned so that the numerical value calculated from the difference between the modeled spectrum and the observed spectrum (standard deviation) stays as small as possible without causing a significant mismatch in relative visibility (i.e., even if we obtained a model that shows a better numerical match in the MW spectrum than in previous model, if this resulted in a significant mismatch in relative visibility, this model is rejected and we roll-back to the previous model). This MW-constraining step slightly alters the match obtained in the HXR fine-tuning, so another small HXR fine-tuning is run again, and so on. We iterate these fine-tunings several times to converge to the unified model that can simultaneously reproduce the observed HXR and MW images and spectra. This workflow, which is based on the framework introduced by \citet{2013SoPh..288..549G}, is illustrated in Figure \ref{f4}.

\section{Model Construction}

\subsection{Constraining in HXR}

As the first step in this modeling workflow, we created  three flux tubes to reproduce HXR emission in three different energy channels, guided by the RHESSI HXR images. We could not reproduce all observed sources with a one or two-loop model because the field line rooted in the 50-75 keV footpoint source locations did not trace out the 6-12 keV source nor cross the 20-35 keV ALT source.

\begin{figure}
\figurenum{5}
\epsscale{.3}
\plotone{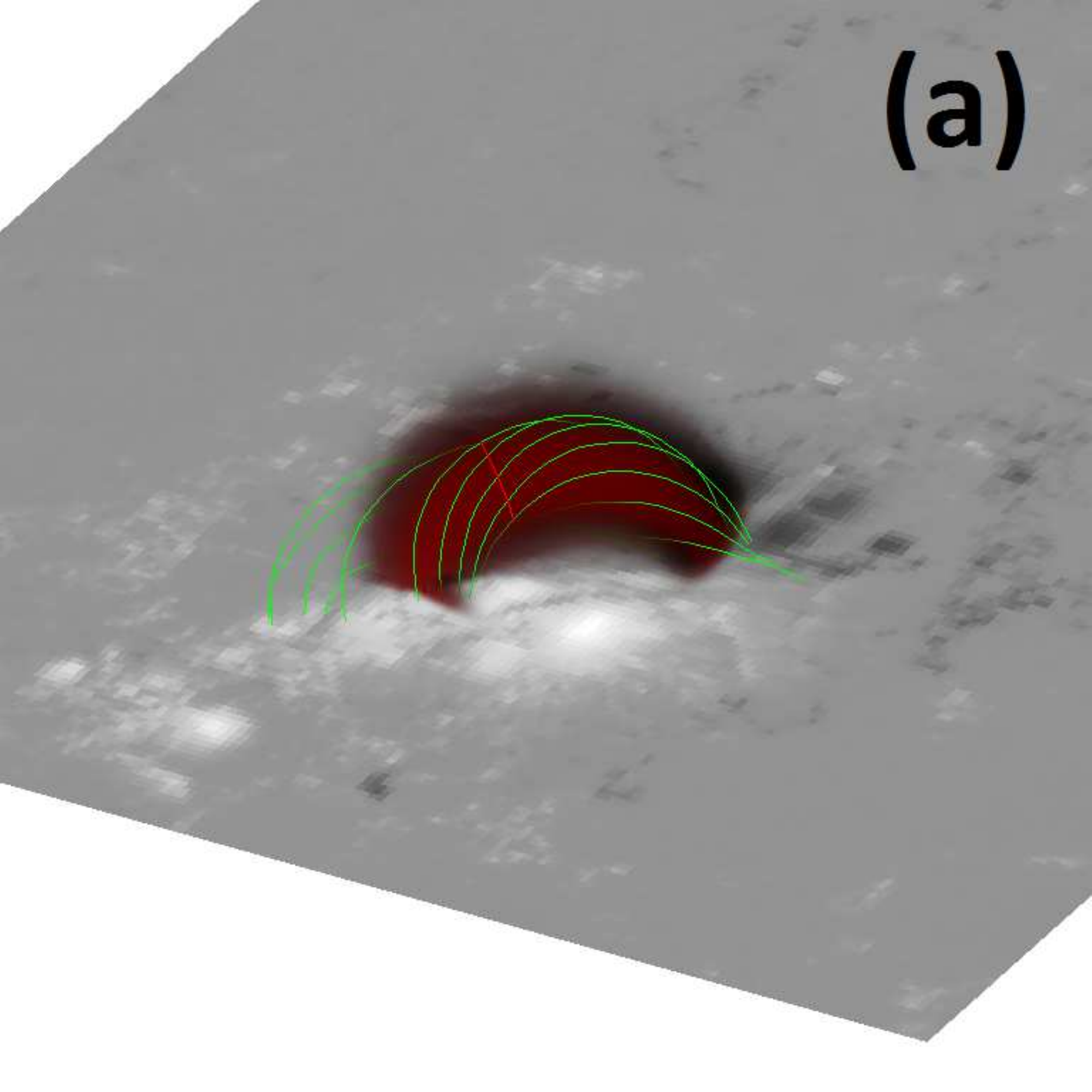}
\plotone{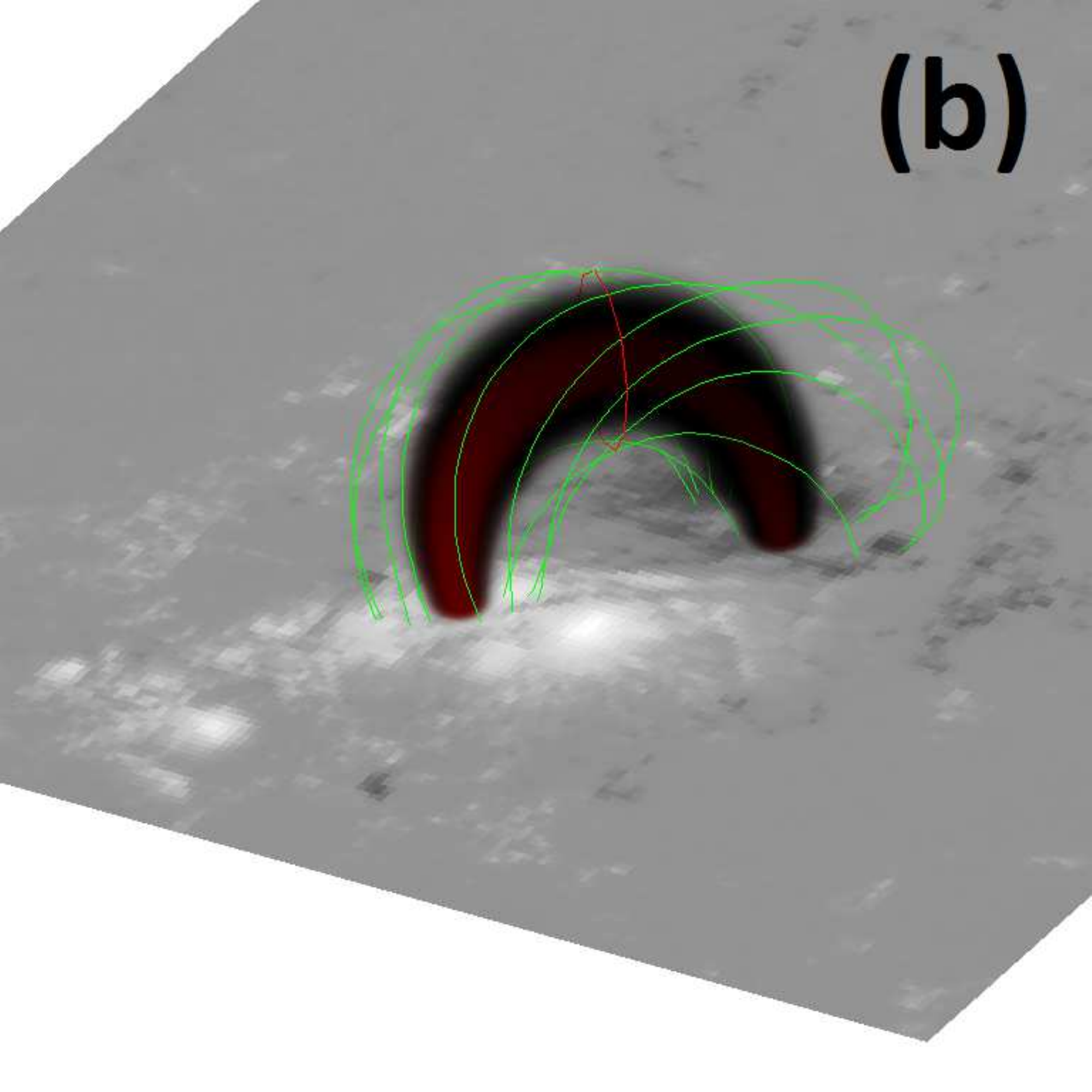}
\plotone{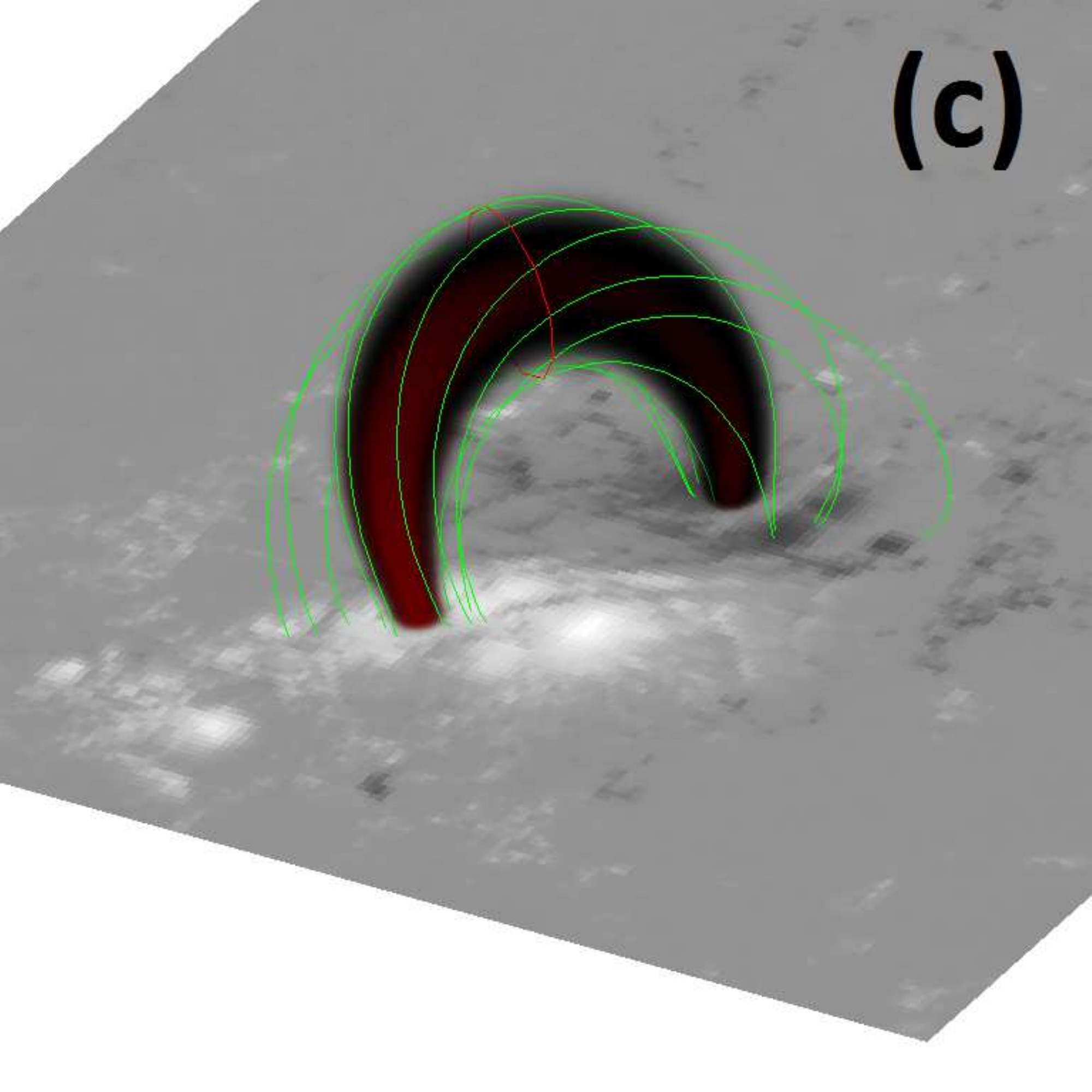}
\plotone{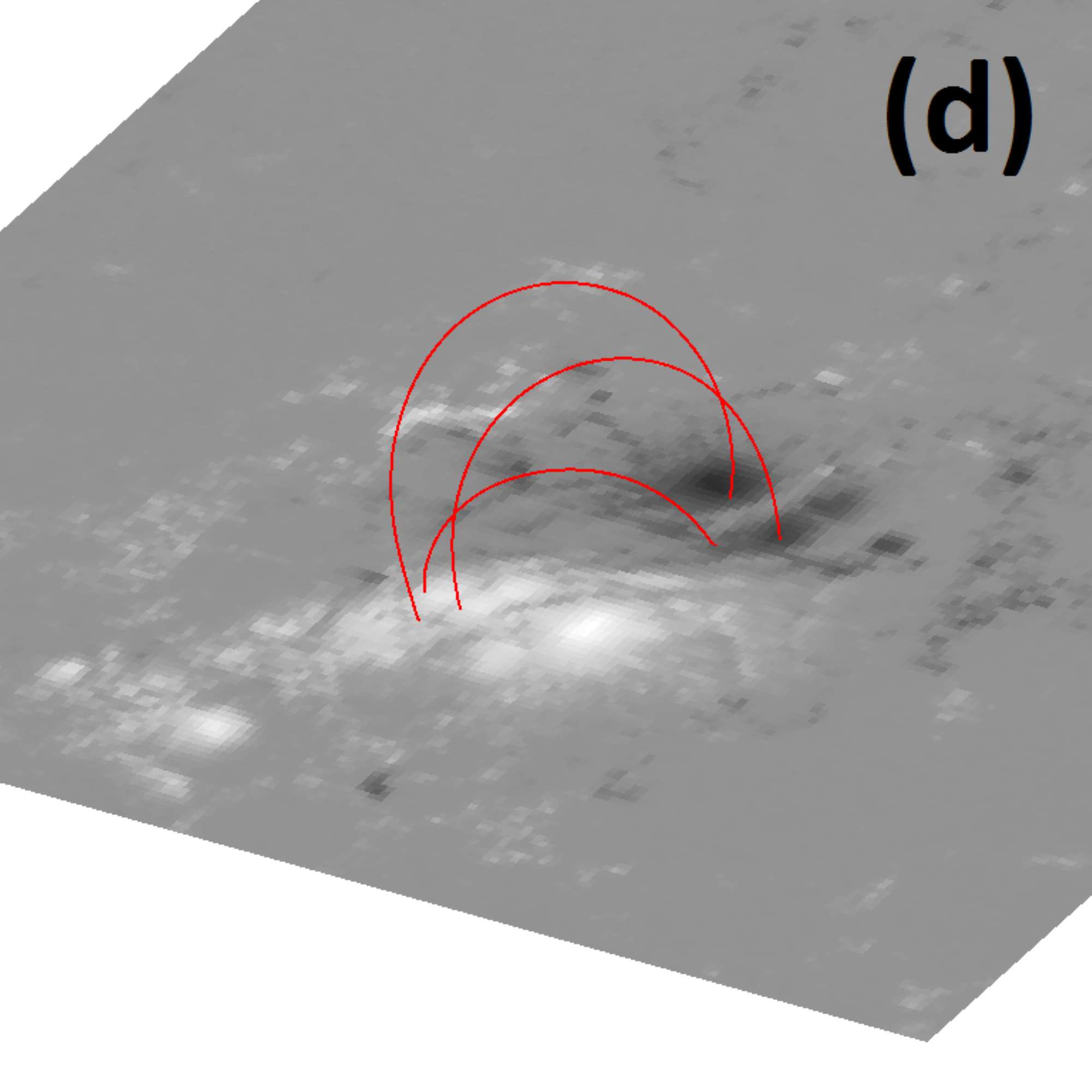}
\plotone{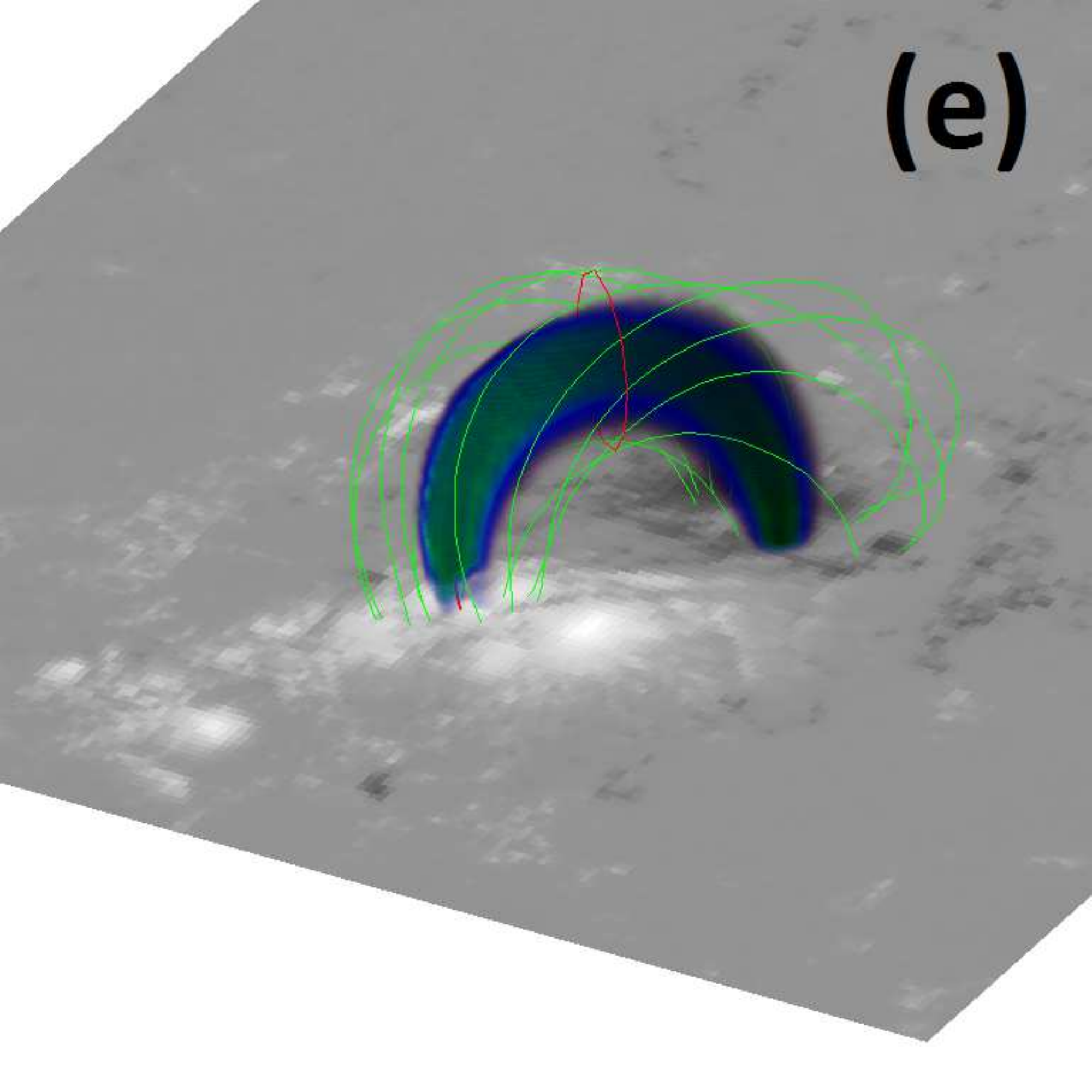}
\plotone{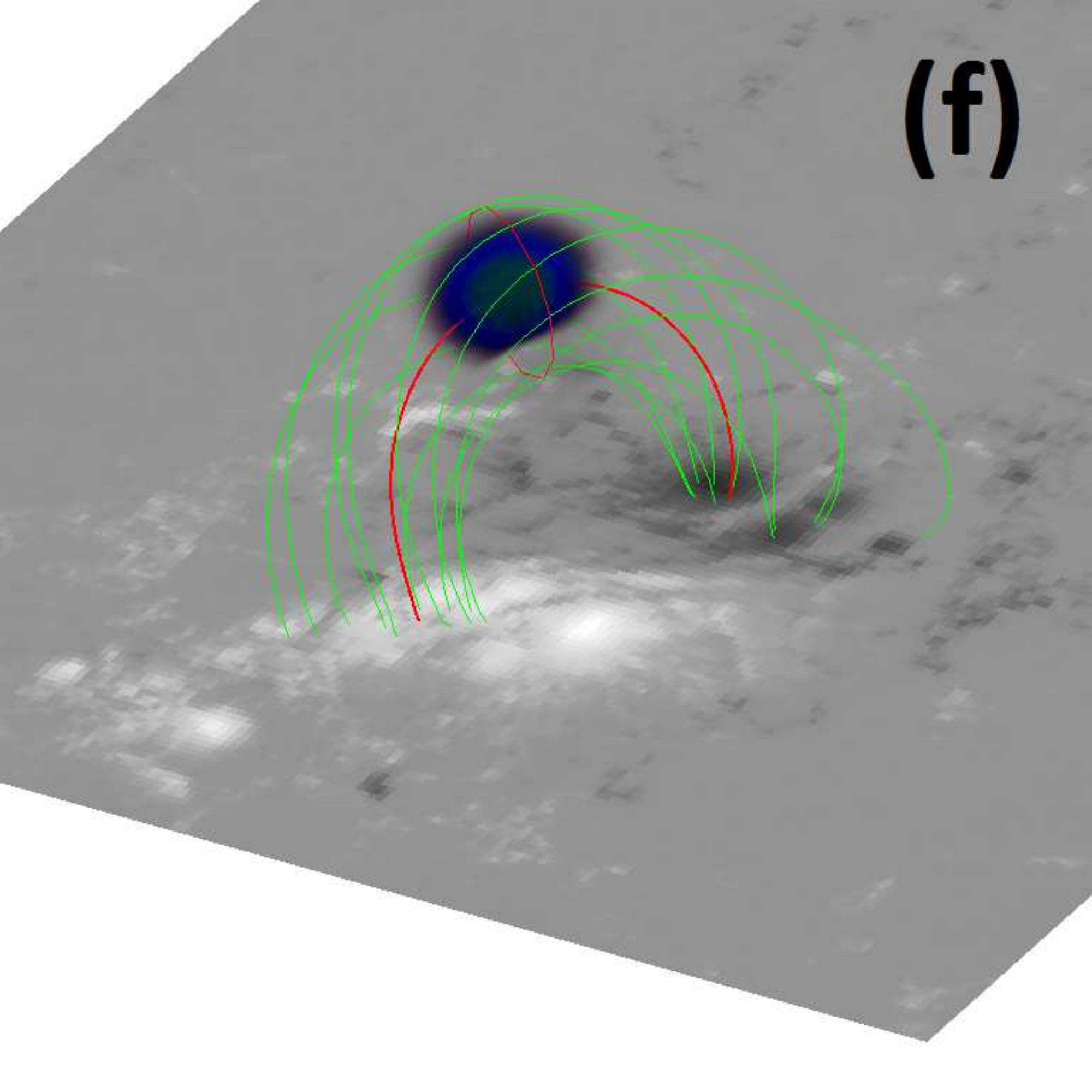}
\caption{The magnetic flux tubes and the corresponding thermal and nonthermal electron populations placed within NLFFF extrapolation cube taken at 18:24 UT, based on the RHESSI image from Figure \ref{f2}(a). (a) Thermal population occupying the flux tube representing the 6-12 keV source, slightly concentrated at the top of the loop. (b,c) Thermal population occupying the flux tube represernting the 50-75 keV double footpoint sources and 20-35 keV ALT source, respectively. (d) The central field lines of three flux tubes shown together within the model. (e) Nonthermal population occupying the flux tube representing the 50-75 keV double footpoint sources. The footpoint will be enhanced in the model HXR image, since a dense chromosphere (not shown) will be included in the final calculation. (f) Nonthermal population occupying the flux tube representing the 20-35 keV ALT HXR source, highly concentrated at the top of the loop. The nonthermal population for the flux tube (a) is not shown because this loop is assigned with zero nonthermal electron density, assuming that it is dominated by thermal electrons. Note that color hues are used only for visual purpose and are scaled individually for each plot. \label{f5}}
\end{figure}

Figure \ref{f5} shows each flux tube and their respective thermal and nonthermal populations. Note that we will be representing the non-ALT 20-35 keV source by the footpoint emissions from the flux tube representing 50-75 keV sources, since they are spatially close to 50-75 keV footpoint sources in the RHESSI image and creating another loop footed at the non-ALT 20-35 keV sources will overcomplicate the model. The first flux tube, in Figure \ref{f5}(a), is the one representing the 6-12 keV source, lying low in the corona with the apex height of $\sim$8,200 km from the photosphere, filled with the thermal population slightly concentrated at the looptop. We note that it was necessary for us to choose such a low-lying structure based on the observed 6-12 keV image; any field lines higher than the selected field line resulted in the misorientation of the model source compared to the observed source. We call this loop ``Thermal-only loop''. For the thermal population, we kept the temperature predicted by OSPEX (20 MK), but altered the vertical density spatial distribution by adding a simple Gaussian-like function to the default function (Appendix Eqn. B4) to match the observed source shape. Then, we fine-tune the density so that the thermal part of the HXR spectrum is well reproduced. As a double check of the model validity we checked that the emission measure calculated from the simulator is consistent with the one predicted from OSPEX. As a result, we found that the thermal population has a density of 1.1 $\times 10^{11}$ cm$^{-3}$ at the bottom and 1.6 $\times 10^{11}$ cm$^{-3}$ at the top of the loop. We assign this flux tube zero nonthermal particle, assuming that this loop is dominated by thermal plasma from chromospheric evaporation - an expected outcome from earlier episodes of particle acceleration.

The second flux tube, Figure \ref{f5}(b,e) (middle column), is the one representing the 50-75 keV double footpoint sources. Both thermal and nonthermal populations are assigned with default spatial distribution functions (essentially uniform for the former and a simple Gaussian-like distribution for the latter) in the loop with the height of $\sim$21,000 km from the photosphere. Note that the HXR emission from this loop will be dominated by footpoints even though the nonthermal electrons are filling the loop, due to a dense chromospheric layer within the magnetic field cube (defined with plasma density $n_{0} = 10^{13}$cm$^{-3}$, $T$ = 3,500K, and the depth of $\sim$2000 km, the default values assigned by the simulator). We call this loop ``Lower loop'', and assign its thermal populations a density of $n_{0} = 5.0 \times 10^{9}$ cm$^{-3}$, which is the default value of the simulator, about an order of magnitude lower than the value assigned for thermal-only loop. For the temperature, we assign 20 MK. These two values are, however, not strictly constrained since the emission measure is solely constrained by the thermal-only loop. We therefore consider the allowable range for these two parameters later.  For the energy distribution, we kept the lower cutoff energy of 22.1 keV from OSPEX, and fine-tuned the density to be $3.0 \times 10^{7}$ cm$^{-3}$ and the low-energy spectral index to be 5.0 in order for the thick-target emission from the footpoint to match the observed spectrum. Futhermore, we assigned a harder spectral index of 3.0 for energy above 200 keV to simultaneously reproduce the negative spectral slope in the high frequency part of the observed MW spectrum by this population.

The third flux tube, Figure \ref{f5}(c) and (f) (right column), is the one representing the 20-35 keV ALT source, with thermal electrons uniformly distributed (Figure \ref{f5}(c)) and nonthermal electrons highly concentrated near the highest point of the loop (Figure \ref{f5}(f)), which is $\sim$27,300 km above the photosphere. We call this loop ``Higher loop''. We assign the thermal population of this loop the same parameters as the lower loop ($T = 20$ MK and $n_{0} = 5.0 \times 10^{9}$ cm$^{-3}$). However, the density of the nonthermal electrons concentrated at the loop-top is set to be much higher than that of the lower loop, 4.5 $\times 10^{8}$ cm$^{-3}$. This high density is required to make the ALT source intensity competitive with the footpoint emissions produced by the lower loop, which are innately bright due to the interaction of their nonthermal particles with the dense chromosphere (thick-target emission). Furthermore, we found that this dense source has to be concentrated energetically as well, since the ALT source does not appear at all in 50-75 keV image. This requires the nonthermal energy distribution to have a double power-law with a softening break of +3.2 at 43 keV.

\begin{figure}
\figurenum{6}
\epsscale{1}
\plotone{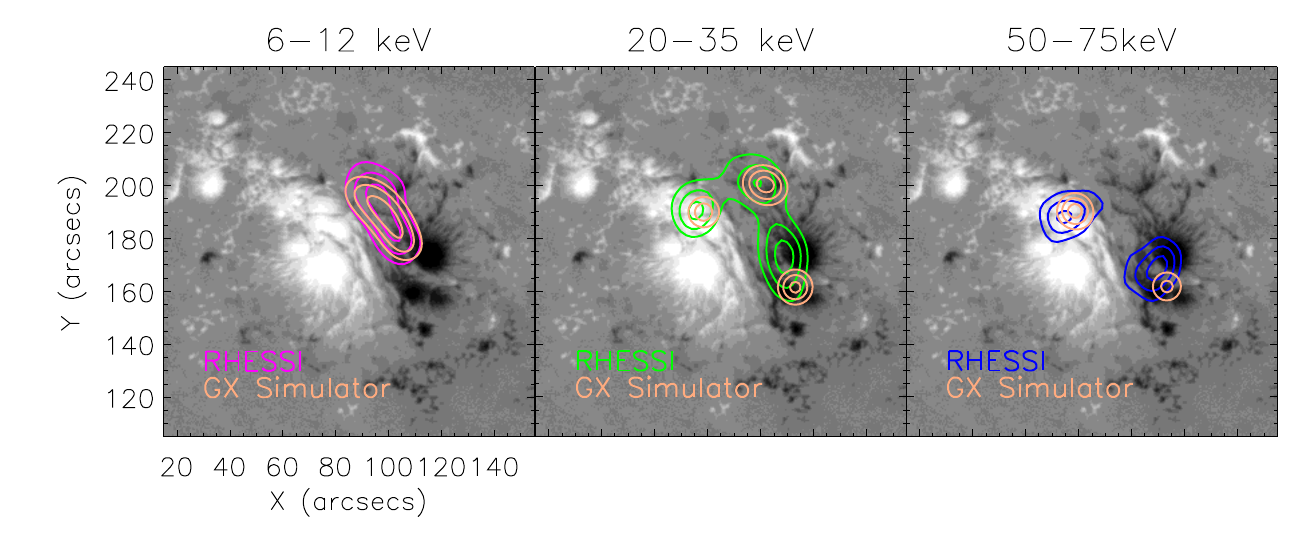}
\caption{The comparison between the observed and the modeled image contours (50, 70, and 90 \%) in three HXR photon energy ranges.  The observed image contours are the same as Figure \ref{f2}(a). The model images are produced as pixelated images by the simulator, so they are further convolved with a Gaussian point-spread-function with the size according to the nominal FWHM resolution of the finest RHESSI collimator used in image reconstruction (collimator 3; 6.79 arcseconds). \label{f6}}
\end{figure}

\begin{figure}
\figurenum{8}
\epsscale{.8}
\plotone{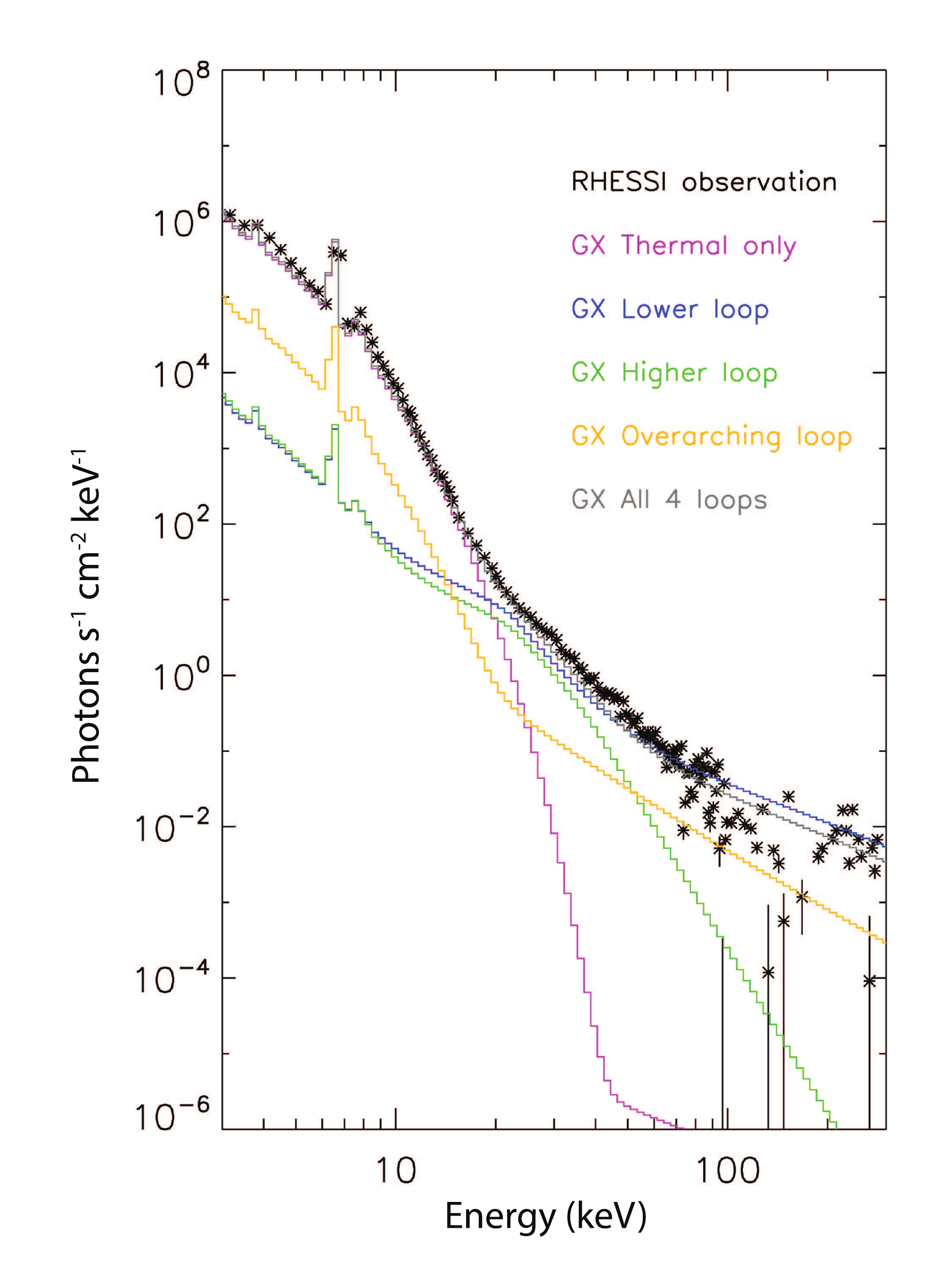}
\caption{The comparison between the observed and the modeled HXR photon spectra, with the breakdown of the contributions from each of four loops in the final model. Grey curves are the emission from the model comprising all four loops in one volume. The slight deficit of the HXR emission from the combined model compared to the emission from the lower loop ($\gtrsim$ 70 keV) is due to the overlapping of some of the loops that interferes with the simulator's voxel ownership implementation (see Section 4.2), but this is not significant since the observed spectrum is statiscally valid only up to 67 keV.\label{f8}}
\end{figure}

Figure \ref{f6} shows the comparison between the observed and the modeled image contours in three HXR photon energy ranges, resulting from the model of these three loops. Figure \ref{f8} shows their contributions (pink, blue, and green), compared to the observed HXR photon spectrum from RHESSI (black). It is evident that the HXR model is complete with the combination of three loops, and our modeled spectrum matches well with the observed spectrum in the statistically meaningful energy range (up to 67 keV).

\subsection{Constraining in MW}

\begin{figure}
\figurenum{7}
\epsscale{1}
\plottwo{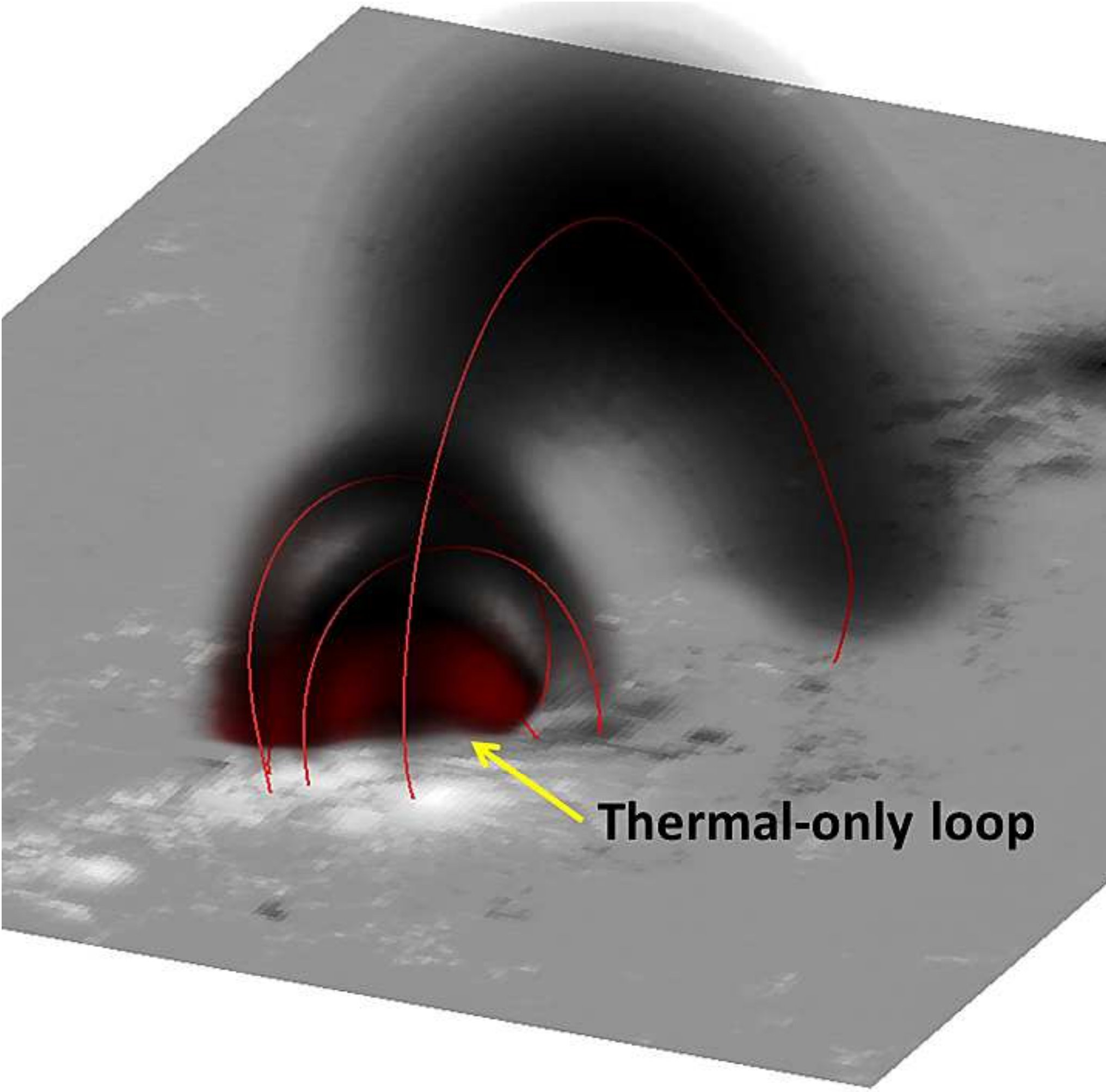}{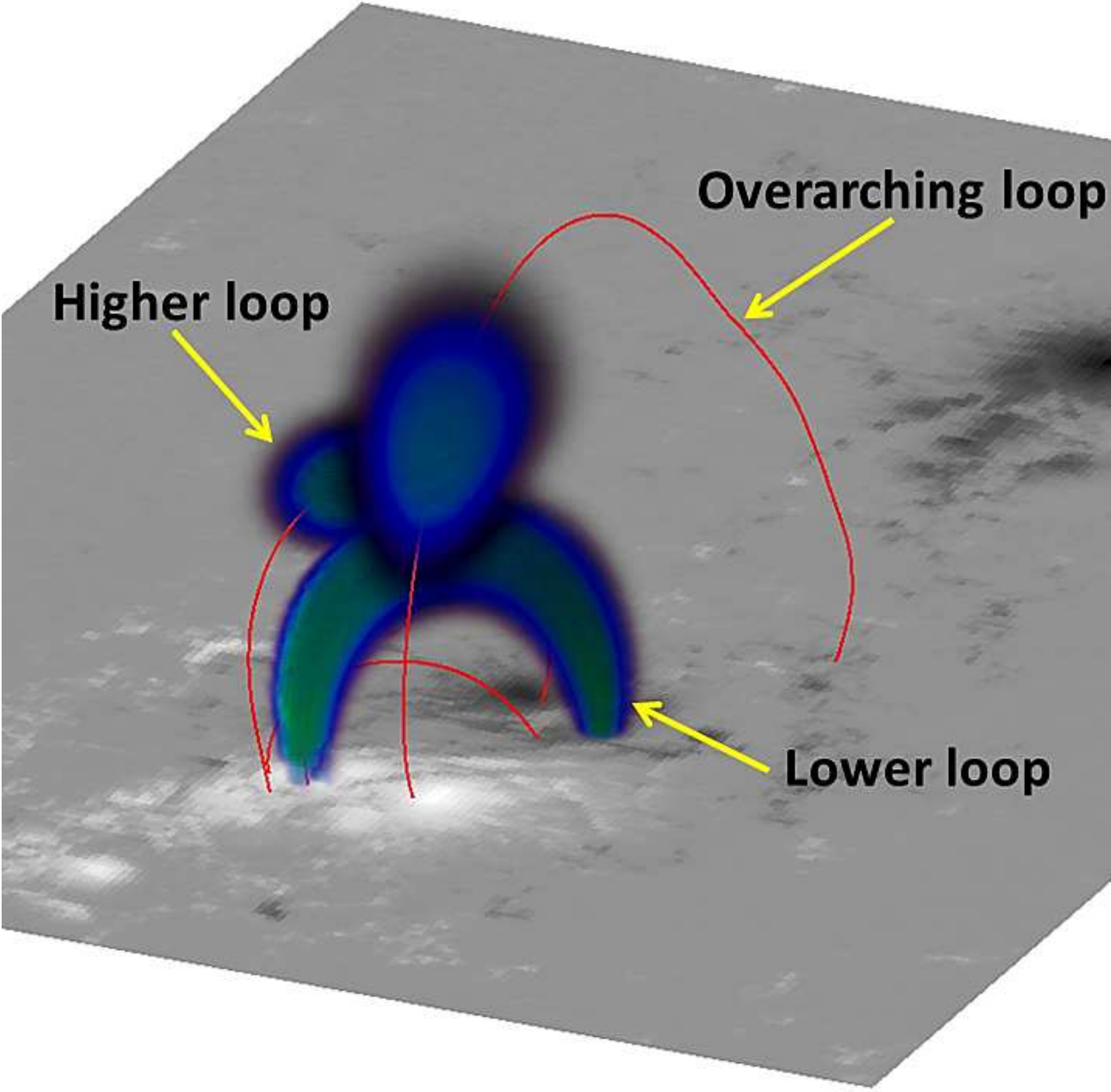}
\caption{The visual representation of thermal (left) and nonthermal (right) electron populations occupying four loops found in the final model. The base image is the HMI photospheric magnetogram taken at 18:24 UT, and the red lines are the central field lines of four loops. The color hues are not in actual density scale. The detailed electron parameters for each named loop are shown in Table 2. \label{f7}}
\end{figure}

\begin{figure}
\figurenum{9}
\includegraphics[scale=.7,angle=90]{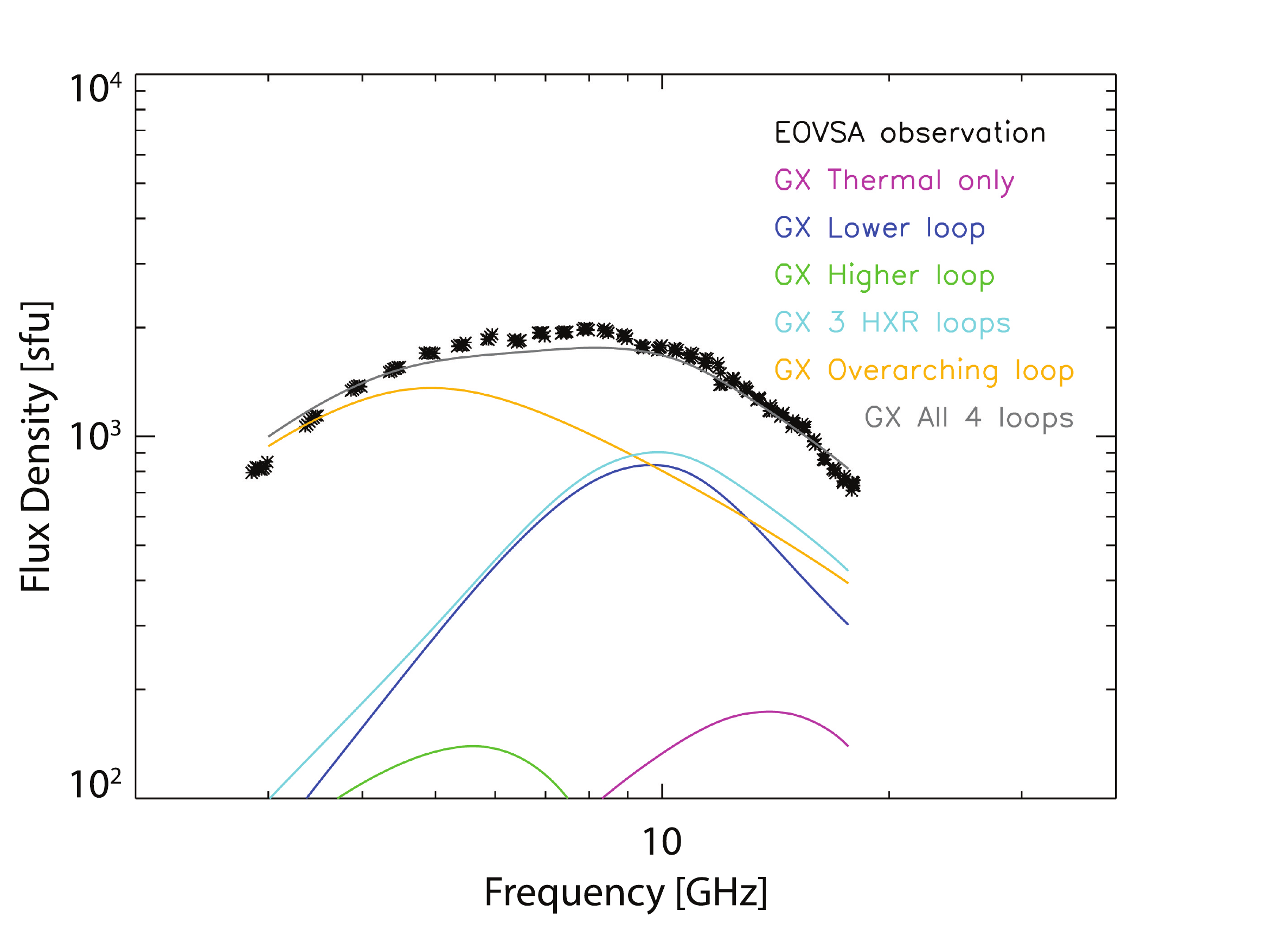}
\caption{The comparison between the observed and the modeled MW total integrated flux density spectra, with the breakdown of the contributions from each of four loops in the final model. Light blue curves are the emission from the model comprising three HXR-constrained loops, and grey curves are the emission from the model comprising all four loops in one volume. \label{f9}}
\end{figure}

Next, we must test this HXR-constrained model in MW. Figure \ref{f9} shows the contributions from the three HXR-producing loops in MW (pink, blue, and green, with light blue showing the sum of three loops), compared to the observed MW total intensity spectrum from EOVSA (black). It is clear that the total MW emission from these three loops does not completely fill the spectrum, especially in the low frequency part. The total area of the HXR-constrained model is too small to reproduce the observed MW emission intensity in this region. Therefore, we considered the existence of another flux tube and its associated electron population that emits mostly in the low frequency MW range, but is ``HXR invisible'', i.e. does not emit significantly in HXR. This is the ``overarching loop" that we investigated. This source cannot be created by a thermal population since the observed MW spectrum's optically-thick flux density requires a high brightness temperature ($>\sim100$ MK), which is unrealistically high for a true thermal temperature. Therefore, this source must be created by a nonthermal population. As mentioned in Section 2.2, the emission over the entire frequency range is produced by GS mechanism. For the GS emission from nonthermal particles to fill the low frequency part of the spectrum, the magnetic field strength has to be relatively weak to lower the peak frequency. Considering this, together with the above-mentioned condition that the area must be larger than the HXR-constrained loops, we created a flux tube that is located above the HXR-emitting loops and has $\sim$10-12 times larger volume than the two other nonthermal loops. Figure \ref{f7} shows our final four-loop model that includes the three HXR-constrained loops and MW-constrained overarching loop. We initially assigned to the overarching loop the identical thermal particle population as the other two loops ($T = 20$ MK and $n_{0} = 5.0 \times 10^{9}$ cm$^{-3}$). Fine-tuning this model loop requires the use of the relative visibility plots from Figure \ref{f3}(a,b), and after several iterations (including the second fine-tuning in HXR, in both images and spectrum), we obtain a nonthermal population that has the density of 6.1 $\times 10^{6}$ cm$^{-3}$ concentrated at an intermediate height in the loop, where the magnetic field is weak. The concentrated density spatial distribution was required by the characteristic source size constraint derived from the observed relative visibilities. The relative location of this population within the flux tube, which is close to the three HXR-emitting loops (Figure \ref{f7}, right), suggests that this population may have been part of the population occupying the HXR-emitting loops but was transported and was accumulated in the region of weaker magnetic field. Its nonthermal energy distribution is found to have a spectral index of 2.5, if we assume a single power-law, and a cutoff energy of 22.1 keV like the other two nonthermal loops. Figure \ref{f3} blue curves shows the $\ln(V_{Rel})$ vs. $B_\lambda^2$ plots from the model. It is evident that the slopes of the model plots are in reasonable agreement with those of the observation in both baselines with different orientation angles, validating that our model is successfully reproducing the observed source size in two different directions. A slight size difference between the model and the observation in the shorter baseline (b) is calculated to be about 4 arcseconds.


The HXR emission contribution from this loop is evidently lower than the other three HXR-constrained loops (Figure \ref{f8}, yellow), and it is so compared to the lower loop because of the difference in the emitting region: chromosphere for the lower loop, and high corona for the overarching loop. The former is thick-target emission while the latter is thin-target emission. Compared to the higher loop, on the other hand, the emission is smaller in the overarching loop simply due to the lower nonthermal particle density; both populations are concentrated within the coronal part of the loop, but the higher loop has $10^{8}$ cm$^{-3}$ nonthermal electrons while the overarching loop has $10^{6}$ cm$^{-3}$ nonthermal electrons. The modeled emission intensity from the overarching loop is lower than the observed emission intensity by an order of magnitude or more for most of the statistically meaningful energy range (less than $\sim$67 keV), and this may be the reason why this source is ``invisible'' in HXR, as the dynamic range of RHESSI is about 10 \citep{2002SoPh..210..287S,2002SoPh..210...61H}. We note that the total emission from all four loops combined (grey curve) is slightly less than the sum of the emission from individual loops (other colors). This is because the model comprising all four loops is a complex system with several loops crossing with each other, and in such cases the simulator is designed to calculate the emission with a physically reasonable approximation that the voxels of higher $n_{0}T$ pressure dominate the contribution in the total emission. In this model, the thermal-only loop is partially intersecting with the lower loop at one of its footpoints, and since the former has higher emission measure (thus higher $n_{0}T$) than the latter and has no nonthermal particles, the emission from the lower loop in the combined model is slightly suppressed in one footpoint. This does not affect our result, however, since the observed spectrum is statistically valid only up to 67 keV. The effect is also confirmed to be negligible in MW.

\subsection{Possible Ranges of The Modeled Parameters}

We now discuss the possible ranges of some of the modeled electron parameters that were not strictly constrained in the model construction process presented above. First, the density and the temperature of the thermal population for the three nonthermal loops (the higher, the lower, and the overarching loops) are not strictly constrained since the emission measure of the entire model is solely constrained by the thermal-only loop. We obtain the ranges for the density and the temperature by fixing one of the two parameters at the values assigned during the model construction and varying the other (the target parameter) until noticeable changes start to appear in the spatially integrated HXR model spectrum. We separately test the overarching loop and the combination of the lower and the higher loops, since their effective volumes are largely different. For the combination of the lower and the higher loops, we find the upper limit of the temperature to be $\sim$50 MK (excess in $\sim$10-20 keV) and that of the density to be $\sim 2 \times 10^{10}$ cm$^{-3}$ (excess in $\sim$20-40 keV). For the overarching loop, we find the upper limit of the temperature to be $\sim 30$ MK (excess in $\sim$10-20 keV) and of the density to be $\sim 2 \times 10^{10}$ cm$^{-3}$ (excess below $\sim$20 keV). We find virtually no lower limit for the two parameters since any emission measure values below the observational constraint is allowed for these loops.

For the nonthermal population, the density of the higher loop and the minimum energy of the overarching loop are not strictly constrained. For the nonthermal density of the higher loop, we must consider its lower limit corresponding to the upper limit for the thermal density, to keep the same level of nonthermal bremsstrahlung from the thin-target coronal source. We find this lower limit to be $\sim 10^{8}$ cm$^{-3}$. The upper limit should be that of the thermal density, $\sim 2 \times 10^{10}$ cm $^{-3}$. For the overarching loop, the electron spectrum was modeled only in the MW range, and since the effective energy of the MW-emitting electrons is above several hundred keV, the spectrum below this energy cannot be strictly constrained. We obtain the more strictly constrained energy range and the number density by testing several different values of cutoff energy and the corresponding normalized number density. As a result, we find that at least 4 $\times 10^{4}$ cm$^{-3}$ nonthermal electrons above E$_{cutoff} \sim 600$ keV are needed to match the low-frequency MW observations.

Lastly, we note that our proposed E$_{break} = 200$ keV for the lower loop nonthermal population is rather arbitrary, since this is above the statistically meaningful energy range of the RHESSI observation (67 keV). To investigate this further, we have eliminated the $\delta_{2}$ above 200 keV from this loop, and evaluated if the high frequency part of the observed MW spectrum can be reproduced solely by the overarching loop. As a result, we found that the overarching loop can reproduce the total intensity spectrum, but cannot solely reproduce the observed relative visibiliy; the source size suggested from the modeled relative visitility becomes too large. We also note that, as shown in Figure \ref{f3}(c), the observed spectra taken at different times during the flare suggest the existence of two components, one above $\sim$ 8 GHz and another below $\sim$ 8 GHz that shows a different temporal variation. Therefore, we strongly believe that the high frequency part of the observed MW emission has to be produced by a flatter electron spectral index in the lower loop (the higher loop cannot produce it due to the steep $\delta_{2}$ required by the HXR observational constraint). Having convinced ourselves of the need for a spectral break in the lower loop, we investigate the allowable range of E$_{break}$ by testing several E$_{break}$ values against the relative visibility calculated from the newly fine-tuned MW model. We find that we can obtain an acceptable relative visibility spectrum with E$_{break}$ up to $\sim$ 220 keV, with slightly increased nonthermal electron density for the overarching loop, $6.7 \times 10^{6}$ cm$^{-3}$, to compensate the deficit caused by such increase in E$_{break}$ in the lower loop. We find the lower limit to be $\sim$ 180 keV, at which the model relative visibility is acceptable but model HXR spectrum starts to show the upward break below 67 keV. For this E$_{break}$ value, slightly lower nonthermal electron density of $5.7 \times 10^{6}$ cm$^{-3}$ is required for the overarching loop, to counterbalance the increase caused by such decrease in E$_{break}$ in the lower loop.

\begin{figure}
\figurenum{10}
\epsscale{1.0}
\plotone{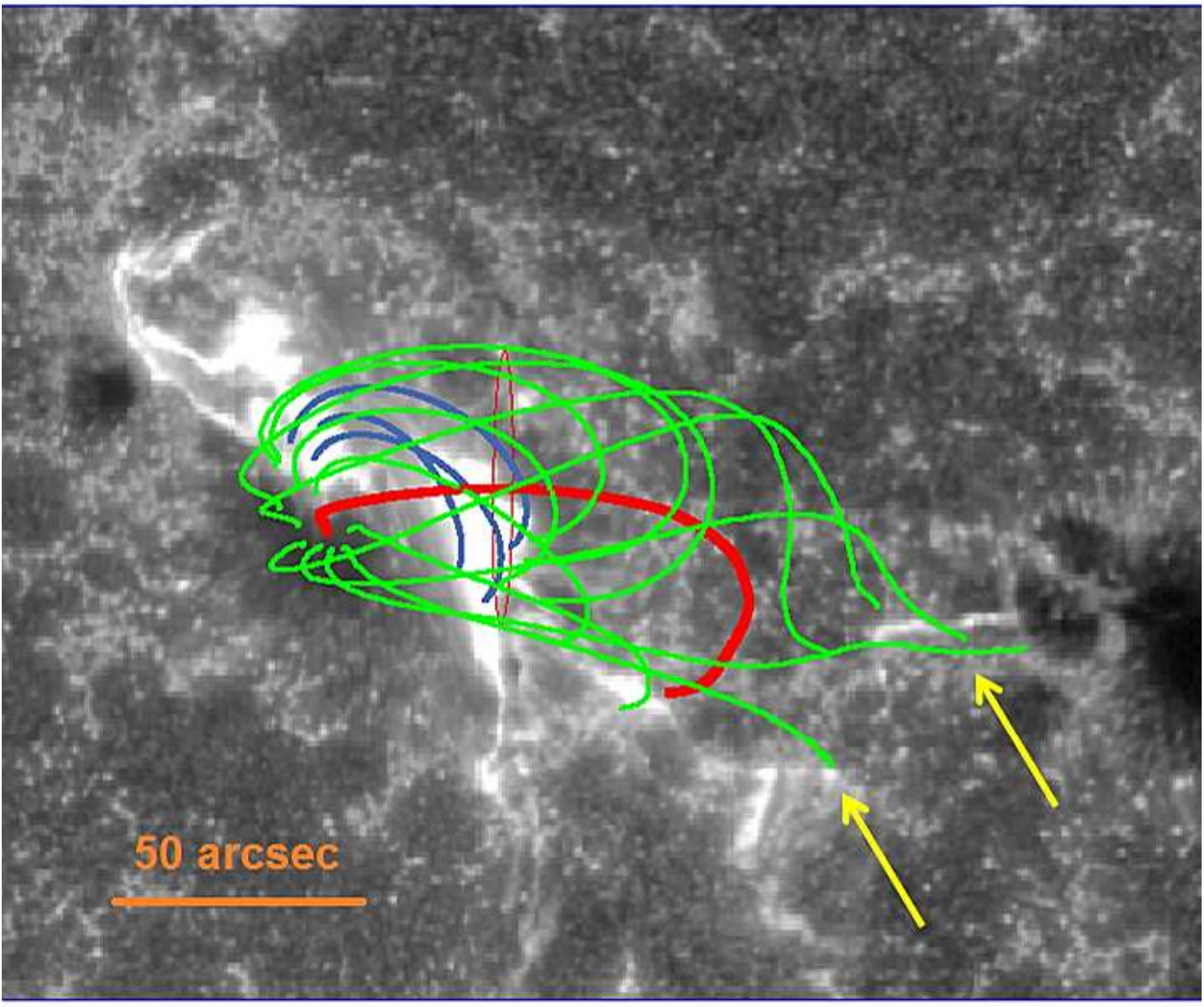}
\caption{The top view of our four loop model overlayed on AIA 1600 $\AA$ image taken at 18:05:28 UT. The blue lines are the central guiding field lines of the three HXR-constrained loops, the red line is the central guiding field line of the overarching loop, and the green lines are the enveloping field lines of the overarching loop. The red thin ellipse in the middle is the top view of the circular cross section of the overarching loop, defining the extent of green field lines. Only the field lines contained in the overarching loop are shown. The western end of this loop seems to match the locations of the remote brightenings indicated by the yellow arrows, which can be interpreted as the precipitation of the nonthermal electrons into the chromosphere on the magnetically weaker side of this loop. \label{f11}}
\end{figure}

Table 2 summarizes the spatial and the energy parameters of each loop and their thermal and nonthermal electron populations, with possible ranges for the unconstrained parameters discussed above shown in red. As stated in the Introduction, since this simulation is a forward-fitting simulation, there could be other ``solutions'' of flux tubes that can equally reproduce the observed HXR and MW emissions (even more than four loops). In this respect, we consider our model to be a ``minimally optimized'' flux tube model, and believe that our methodology, which starts from the construction of the model based on the HXR observation, was the best approach in obtaining a simultaneous fit to the observational constraints that were available for this event. Also, we coincidentally found several remote brightenings in an AIA 1600 $\AA$ channel image taken at the modeling time that correspond to the western footpoints of the overarching loop in our model, as seen in Figure \ref{f11} (yellow arrows). This correspondence could be interpreted as the signature of the precipitation of the nonthermal particles in the overarching loop into the chromosphere. The fact that the brightenings appear only at the western end of the loop may be due to the large difference in the magnetic field strength (thus the mirror ratio) at the two ends: they are $\sim$2,000 G and $\sim$1,000 G at the eastern and western end, respectively, and only the magnetically weaker end is allowing the particles to precipitate through. The fact that the locations of these remote brightenings closely match with the furthest end of the model loop strongly supports that the size and the extent of our overarching loop is correctly representing the actual flaring loop. For the possible variation in the HXR-constrained model, we briefly consider how our model may vary if we interpret that the 20-35 keV ALT source is a thermal source, as mentioned in Section 2.2. We find that modeling the 20-35 keV ALT source with $n_{0} \sim 10^{10}$ cm$^{-3}$ and $T \sim 60$ MK can equally produce the observed HXR images and spectrum without violating observed emission measure. We cannot, however, currently test the validity of this model since this model will require setting two different temperature values within one loop.

\section{Discussion and Conclusions}

Although a number of solar flares have been analyzed in 3D using the GX Simulator \citep{2017ApJ...845..135F,2016ApJ...822...71F,2016ApJ...816...62F}, all previous studies relied on Potential Field or Linear Force-Free Field extrapolation. We present here the first flare model that contains multiple flaring loops inside a single NLFFF data cube. Based on the results of our modeling, we draw the following conclusions about the spatial and the energy distributions of the energetic electrons producing HXR and MW emissions at the end of the impulsive phase of 2015-06-22 M6.5 flare.

First, based on the observed HXR emission sources, the magnetic field configuration that best represents the flaring loop geometry in our study was found to be the post-reconnection loop configuration. We compared the field line connectivity of the NLFFF extrapolation cube created near the modeling time of 18:05:32 UT, and found that field lines contained in the cube created at the earlier time did not have the desired connectivity for our observed HXR sources due to its sheared overall field line geometry. It is evident that our modeled peak has not started at 18:00 UT yet (see Figure \ref{f1}(c)), so it is reasonable to think that the shear reduced due to the reconnection event responsible for our modeled peak. The chosen cube at 18:24 UT should contain more post-reconnection loops, and although the field model with finer time cadence considering the dynamics of the flaring loop may still improve the accuracy of the model, our results show that this post-reconnection cube can reproduce the relative locations of the observed HXR sources and the characteristic size of the observed MW source well.

Second, the low frequency part of the MW spectrum is dominated by the emission from a ``HXR invisible'' source containing a non-negligible number of nonthermal electrons in a relatively large volume with relatively weak magnetic field. The nonthermal particle population in the overarching loop fills a volume $\sim$10-12 times larger than the other two nonthermal loops. The total number of nonthermal electrons contained in this loop is calculated to be $\sim 10^{34}$, which is in the same order of magnitude as that calculated from the lower loop, $\sim 5 \times 10^{34}$, the main contributor of both HXR and MW high-frequency emission. We conclude that the primary reason that this MW low-frequency source is ``HXR invisible'' although it contains up to $6.1 \times 10^{6}$ cm$^{-3}$ nonthermal electrons is because the emitting electrons are trapped in the high corona, and in HXR, the emission from this source is overcome by the bright thick-target footpoint emission. This population has an interesting nonthermal energy spectrum with the spectral index of 2.5, the hardest of all three nonthermal populations. We interpret this  as a result of particle accumulation and trapping at that location above the main loops, caused by some transport process underway throughout the impulsive phase. Such an interpretation is also reasonable if one notices that the low frequency part of the MW spectrum grew over several minutes only toward the end of the impulsive phase, as evident in the 4.43 GHz lightcurve (red) in Figure \ref{f1}(c). Our results also showed that the parameters of this overarching loop are essentially insensitive to the possible variation in the model parameters of the lower loop, which contributes to the high-frequency part of the spectrum, because the lower loop is too small to provide the GX flux level at low frequencies \citep[see Figure \ref{f9} blue curve, also][Eq. 1]{2017ApJ...845..135F}. In other words, the deficit in the low frequency part of the MW spectrum will be present as long as we confine the main HXR loops compactly at lower heights based on the observed HXR images.

Third, the overall geometry and the locations of the electron populations in the three HXR-emitting loops in our model is consistent the standard flare scenario. The  thermal-only loop can be interpreted as a result of chromospheric evaporation in response to earlier electron acceleration episodes; dense ($10^{11}$ cm$^{-3}$) thermal plasma concentrated toward the looptop. It is interesting that this loop was found to lie relatively low, and this may be a result of our choice of the NLFFF magnetic field model from a time (18:24 UT) that is later than 18:05:32 UT, which contains more post-reconnection loops. However, considering that our modeling time is already near the end of the impulsive phase when most of the loops have reconnected and became low-lying post-reconnection loops, we consider that our modeled geometry is correctly representing the actual geometry of the flare loops. The lower loop population is the major contributor in the ``HXR-visible'' nonthermal population, and can be considered as the traditional ``common'' population of nonthermal electrons, if its energy spectrum has a broken power-law distribution. The low-energy end of this population produces HXR thick-target emission from the chromosphere with a soft spectrum, and the high-energy end of this population produces high-frequency MW emission from the corona with a hard spectrum. The break energy of this population is modeled to be in the range $\sim180$-220 keV. We cannot test if the spectral hardening of -2 at these energies was already present coming out of the acceleration region, or because some of the lower energy population moved to the higher energy population via second-stage acceleration that preferentially accelerates higher energy electrons during the propagation and the trapping, or both.

In summary, our results show that our three-dimensional forward-fit modeling of the flare HXR and MW emission can reveal the properties of the nonthermal particles in the flare in much greater quantitative detail than those obtained by conventional means, both within and outside of the standard flare model. We would like to emphasize the importance of our finding, the existence of the ``HXR invisible'' nonthermal particles that can only be investigated through the properties of the MW low frequency emission. This suggestion is not new \citep[e.g.,][]{1994SoPh..152..409L,2016ApJ...822...71F,2017ApJ...845..135F}, but has been largely neglected in the standard flare model because the focus of the flare-accelerated electrons has been mainly in the HXR range (and their counterpart in MW high frequency range) where the bulk of their energy is deposited. Even though these low frequency MW-emitting electrons are still at the ``tail'' of the electron number distribution, their high energy and the trapped condition may make them an energetically important player in the overall flare energetic scenario, even after the impulsive phase. For instance, these trapped high-energy electrons \citep[cf.][]{2016ApJ...822...71F,2017ApJ...845..135F} may escape directly or be further accelerated by CME shocks and become Solar Energetic Particles. Our modeling also stimulates the further investigation into the possible spectral break in the population emitting in both HXR and MW range. Our model for this particular flare seems to support the existence of the break, but other flares may not show such a break. Running this type of modeling for many other flares could lead to the findings of the properties of the flare that may or may not result in a break. Such information should enable us to discriminate among the competing models of flare particle acceleration, trapping, and escape. 

\acknowledgements
This work was supported in part by NSF grants AST-1312802, AGS-1348513, AGS-1408703, AGS-1262772, AST-1615807, and NASA grants NNX14AC87G, NNX13AG13G, and NNX16AF72G, NNX14AK66G, 80NSSC18K0015, 80NSSC17K0016 to New Jersey Institute of Technology. We thank SDO and RHESSI teams for the data used in this publication. We thank E. P. Kontar for his invaluable timely contribution to the implementation of the thick-target model to GX Simulator. We thank A. Tsvetkova for placing our model on the website at \url{http://www.ioffe.ru/LEA/SF_AR/models/3dmodels.html}.

\appendix
\section{Appendix: Derivation of one-dimensional relative visibilities for a simple Gaussian source}

Visibility on a particular baseline is defined as the Fourier transform of the sky brightness distribution. Let us assume a simple Gaussian source flux function:

\begin{equation}
S(x) = a e^{\frac{-(x-x_{0})^2}{\alpha^{2}}}
\end{equation}

\noindent
where $S(x)$ is the flux intensity as a function of spatial coordinate $x$, $a$ is a unit amplitude at $x = x_{0}$, and $\alpha$ is the 1/e width of the unit amplitude (half power beam width). Then the visibility as a function of spatial frequency $s$ ($x/\lambda$) is

\begin{equation}
V(s) = \int_{-\infty}^{\infty} S(x) e^{-2 \pi isx} dx
= a e^{-x_{0}^2 / \alpha^2} e^{\left(\frac{x_{0}}{\alpha^2} - \pi is\right)^2} \alpha^2 \int_{-\infty}^{\infty} e^{-\left[\frac{x}{\alpha} - \alpha\left(\frac{x_{0}}{\alpha^2} - \pi is\right)\right]^2}
= a \sqrt \pi \alpha e^{- \pi^{2} s^{2} \alpha^{2}} e^{-2 \pi ix_{0}s}
\end{equation}

\noindent
Relative visibility is defined as the visibility divided by the total power ($s\rightarrow0$),

\begin{equation}
V_{Rel} = \frac{V(s)}{V(0)}
= e^{-\pi^2 s^2 \alpha^2}
\end{equation}

\noindent
where $V_{Rel}$ is a unit amplitude ($x_{0} = 0$). Plugging in the definition of spatial frequency $s$ and the conversion factor between $\alpha$ and the FWHM of the source, $d$,

\begin{displaymath}
s = \frac{1}{\theta} rad^{-1} \sim B_{\lambda} rad^{-1}
= \frac{B_{cm} f_{GHz}}{30} rad^{-1}
= -8.351 \times 10^{-11} B_{\lambda}^2 arcsec^{-1}
= 1.62 \times 10^{-7} B_{cm} f_{GHz} arcsec^{-1}
\end{displaymath}

\begin{displaymath}
\alpha \sim0.6d
\end{displaymath}

\noindent
(A3) becomes

\begin{equation}
V_{Rel} = e^{-8.351 \times 10^{-11} B_{\lambda}^2 d^2}
= e^{-9.325 \times 10^{-14} B_{cm}^2 d^2 f_{GHz}^2}
\end{equation}

\noindent
where $\theta$ is the fringe spacing, $B_{\lambda}$ is the projected baseline length in number of wavelength, $B_{cm}$ is the projected baseline length in cm, $f_{GHz}$ is the observing frequency in GHz, and $d$ is in arcsec. Note that $V_{Rel}$ here is a unit amplitude ($x_{0} = 0$). In practice, $V(s)$ is the cross-correlated amplitude from a particualr baseline and $V(0)$ is the geometric mean of the total power from the two antennas on that baseline. That is,

\begin{equation}
V_{Rel} = \frac{a_{ij}}{\sqrt{{a_{ii}a_{jj}}}}
\end{equation}

\noindent
where $a_{ij}$ is the cross-correlated amplitudes from the baseline consisting of antenna $i$ and $j$, and $a_{ii}$ and $a_{jj}$ are the auto-correlated total power amplitudes from antenna $i$ and $j$, respectively. Combining (A4) and (A5), we have

\begin{equation}
V_{Rel} = \frac{a_{ij}}{\sqrt{{a_{ii}a_{jj}}}}
= e^{-8.351 \times 10^{-11} B_{\lambda}^2 d^2}
= e^{-9.325 \times 10^{-14} B_{cm}^2 d^2 f_{GHz}^2}
\end{equation}

\noindent
or

\begin{equation}
\ln(V_{Rel}) = \ln\left(\frac{a_{ij}}{\sqrt{{a_{ii}a_{jj}}}}\right)
= -8.393 \times 10^{-11} B_{\lambda}^2 d^2
= -9.325 \times 10^{-14} B_{cm}^2 d^2 f_{GHz}^2
\end{equation}

\section{Appendix: List of spatial distribution functions used for this study}

For the spatial distribution of the electron population modeled in this study, we used the default functions provided by the GX Simulator, presented hereafter, and also available with more details in \citet{2015ApJ...799..236N}.

\textbf{Thermal population:}

The default distribution function for thermal population is a product of a normalization factor (density, $n_{0}$), a unit-amplitude radial distribution across the flux tube ($n_{r}$), and a unit-amplitude vertical distribution ($n_{z}$):

\begin{equation}
n_{th}(x,y,z) = n_{0}n_{r}(x/a,y/b)n_{z}[z(s)/R]
\end{equation}

\noindent
where

\begin{equation}
n_{r}(x,y) = \exp\left[-\left(p_{0}\frac{x}{a}\right)^2 - \left(p_{1}\frac{y}{b}\right)^2\right]
\end{equation}

\noindent
where $x$ and $y$ are the cross-sectional coordinate of the flux tube  normalized by the ellipse semi-axis of the cross-section $a$ and $b$, and $z(s)$ is the vertical coordinate intersecting the longitudinal coordinate $s$ of the flux tube. The vertical density distribution, $n_{z}$ is a simple hydrostatic formula adapted from $\S$3.1 of \citep{2004psci.book.....A}:

\begin{equation}
n(z) = \exp\left[-\frac{z(s)/R}{6.7576 \times 10^{-8}T_{0}}\right].
\end{equation}

\noindent
where $R$ is the solar radius and $T_{0}$ is the temperature of the population. The parameters $p_{0}$ and $p_{1}$ determine the radial extent of the population across the flux tube (the smaller the parameter, the broader the extent), while $T_{0}$ determines the vertical distribution of the population. 

Note that, for thermal-only loop in this study, we used a customized $n(z)$ to match the observed HXR 6-12 keV source shape. The equation is the sum of the default function (B3) and a simple Gaussian-like function used for nonthermal population (B6, below) multiplied by a factor of 1.5:

\begin{equation}
n(z) = \exp\left[-\frac{z(s)/R}{6.7576 \times 10^{-8}T_{0}}\right] + 1.5\exp\Bigl\{-\left[q_{0}\left(\frac{s-s_{0}}{l} + q_{2}\right)\right]^2\Bigr\}
\end{equation}

\noindent
where $q_{0}$ and $q_{2}$ determines the longitudinal distribution (see below).

\textbf{Nonthermal population:}

The default distribution function for the nonthermal population is a product of a normalization factor (density, $n_{b}$), a unit-amplitude radial distribution across the flux tube ($n_{r}$, Eqn. (B2) above), and a unit-amplitude longitudinal distribution along the central field line ($n_{s}$):

\begin{equation}
n_{nth}(x,y,z) = n_{b}n_{r}(x/a,y/b)n_{s}(s/l)
\end{equation}

\noindent
where $a$, $b$, and $s$ are defined as above, and $l$ is the length of the central field line of the flux tube. The longitudinal function is a simple Gaussian-like function of the form:

\begin{equation}
n_{s}(s) = \exp\Bigr\{-\left[q_{0}\left(\frac{s-s_{0}}{l} + q_{2}\right)\right]^2\Bigr\}
\end{equation}

\noindent
where $s_{0}$ is the reference longitudinal point along the central field line. The parameter $q_{0}$ determines the longitudinal distribution of the population along the flux tube while $q_{2}$ determines the location of the peak of such distribution.

\bibliographystyle{apj}
\bibliography{bibliography}

\begin{thebibliography}{39}
\expandafter\ifx\csname natexlab\endcsname\relax\def\natexlab#1{#1}\fi

\bibitem[{{Aschwanden}(2004)}]{2004psci.book.....A}
{Aschwanden}, M.~J. 2004, {Physics of the Solar Corona. An Introduction}
  (Praxis Publishing Ltd)

\bibitem[{{Bai} \& {Ramaty}(1978)}]{1978ApJ...219..705B}
{Bai}, T., \& {Ramaty}, R. 1978, \apj, 219, 705

\bibitem[{{Brown}(1971)}]{1971SoPh...18..489B}
{Brown}, J.~C. 1971, \solphys, 18, 489

\bibitem[{{Dennis}(1988)}]{1988SoPh..118...49D}
{Dennis}, B.~R. 1988, \solphys, 118, 49

\bibitem[{{Fleishman} {et~al.}(2017{\natexlab{a}}){Fleishman}, {Anfinogentov},
  {Loukitcheva}, {Mysh'yakov}, \& {Stupishin}}]{2017ApJ...839...30F}
{Fleishman}, G.~D., {Anfinogentov}, S., {Loukitcheva}, M., {Mysh'yakov}, I., \&
  {Stupishin}, A. 2017{\natexlab{a}}, \apj, 839, 30

\bibitem[{{Fleishman} \& {Kuznetsov}(2010)}]{2010ApJ...721.1127F}
{Fleishman}, G.~D., \& {Kuznetsov}, A.~A. 2010, \apj, 721, 1127

\bibitem[{{Fleishman} {et~al.}(2017{\natexlab{b}}){Fleishman}, {Nita}, \&
  {Gary}}]{2017ApJ...845..135F}
{Fleishman}, G.~D., {Nita}, G.~M., \& {Gary}, D.~E. 2017{\natexlab{b}}, \apj,
  845, 135

\bibitem[{{Fleishman} {et~al.}(2016{\natexlab{a}}){Fleishman}, {Pal'shin},
  {Meshalkina}, {Lysenko}, {Kashapova}, \& {Altyntsev}}]{2016ApJ...822...71F}
{Fleishman}, G.~D., {Pal'shin}, V.~D., {Meshalkina}, N., {Lysenko}, A.~L.,
  {Kashapova}, L.~K., \& {Altyntsev}, A.~T. 2016{\natexlab{a}}, \apj, 822, 71

\bibitem[{{Fleishman} {et~al.}(2016{\natexlab{b}}){Fleishman}, {Xu}, {Nita}, \&
  {Gary}}]{2016ApJ...816...62F}
{Fleishman}, G.~D., {Xu}, Y., {Nita}, G.~N., \& {Gary}, D.~E.
  2016{\natexlab{b}}, \apj, 816, 62

\bibitem[{{Gary} {et~al.}(2013){Gary}, {Fleishman}, \&
  {Nita}}]{2013SoPh..288..549G}
{Gary}, D.~E., {Fleishman}, G.~D., \& {Nita}, G.~M. 2013, \solphys, 288, 549

\bibitem[{{Gary} \& {Hurford}(1989)}]{1989ApJ...339.1115G}
{Gary}, D.~E., \& {Hurford}, G.~J. 1989, \apj, 339, 1115

\bibitem[{{Haug}(1997)}]{1997A&A...326..417H}
{Haug}, E. 1997, \aap, 326, 417

\bibitem[{{Hurford} {et~al.}(2002){Hurford}, {Schmahl}, {Schwartz}, {Conway},
  {Aschwanden}, {Csillaghy}, {Dennis}, {Johns-Krull}, {Krucker}, {Lin},
  {McTiernan}, {Metcalf}, {Sato}, \& {Smith}}]{2002SoPh..210...61H}
{Hurford}, G.~J., {et~al.} 2002, \solphys, 210, 61

\bibitem[{{Jing} {et~al.}(2016){Jing}, {Xu}, {Cao}, {Liu}, {Gary}, \&
  {Wang}}]{2016NatSR...624319J}
{Jing}, J., {Xu}, Y., {Cao}, W., {Liu}, C., {Gary}, D., \& {Wang}, H. 2016,
  Scientific Reports, 6, 24319

\bibitem[{{Kundu} {et~al.}(2009){Kundu}, {Grechnev}, {White}, {Schmahl},
  {Meshalkina}, \& {Kashapova}}]{2009SoPh..260..135K}
{Kundu}, M.~R., {Grechnev}, V.~V., {White}, S.~M., {Schmahl}, E.~J.,
  {Meshalkina}, N.~S., \& {Kashapova}, L.~K. 2009, \solphys, 260, 135

\bibitem[{{Kundu} {et~al.}(1994){Kundu}, {White}, {Gopalswamy}, \&
  {Lim}}]{1994ApJS...90..599K}
{Kundu}, M.~R., {White}, S.~M., {Gopalswamy}, N., \& {Lim}, J. 1994, \apjs, 90,
  599

\bibitem[{{Lee} {et~al.}(1994){Lee}, {Gary}, \& {Zirin}}]{1994SoPh..152..409L}
{Lee}, J.~W., {Gary}, D.~E., \& {Zirin}, H. 1994, \solphys, 152, 409

\bibitem[{{Lemen} {et~al.}(2012){Lemen}, {Title}, {Akin}, {Boerner}, {Chou},
  {Drake}, {Duncan}, {Edwards}, {Friedlaender}, {Heyman}, {Hurlburt}, {Katz},
  {Kushner}, {Levay}, {Lindgren}, {Mathur}, {McFeaters}, {Mitchell}, {Rehse},
  {Schrijver}, {Springer}, {Stern}, {Tarbell}, {Wuelser}, {Wolfson}, {Yanari},
  {Bookbinder}, {Cheimets}, {Caldwell}, {Deluca}, {Gates}, {Golub}, {Park},
  {Podgorski}, {Bush}, {Scherrer}, {Gummin}, {Smith}, {Auker}, {Jerram},
  {Pool}, {Soufli}, {Windt}, {Beardsley}, {Clapp}, {Lang}, \&
  {Waltham}}]{2012SoPh..275...17L}
{Lemen}, J.~R., {et~al.} 2012, \solphys, 275, 17

\bibitem[{{Lin} {et~al.}(2003){Lin}, {Krucker}, {Hurford}, {Smith}, {Hudson},
  {Holman}, {Schwartz}, {Dennis}, {Share}, {Murphy}, {Emslie}, {Johns-Krull},
  \& {Vilmer}}]{2003ApJ...595L..69L}
{Lin}, R.~P., {et~al.} 2003, \apjl, 595, L69

\bibitem[{{Liu} {et~al.}(2016){Liu}, {Xu}, {Cao}, {Deng}, {Lee}, {Hudson},
  {Gary}, {Wang}, {Jing}, \& {Wang}}]{2016NatCo...713104L}
{Liu}, C., {et~al.} 2016, Nature Communications, 7, 13104

\bibitem[{{Masuda} {et~al.}(1994){Masuda}, {Kosugi}, {Hara}, {Tsuneta}, \&
  {Ogawara}}]{1994Natur.371..495M}
{Masuda}, S., {Kosugi}, T., {Hara}, H., {Tsuneta}, S., \& {Ogawara}, Y. 1994,
  \nat, 371, 495

\bibitem[{{Nita} {et~al.}(2015){Nita}, {Fleishman}, {Kuznetsov}, {Kontar}, \&
  {Gary}}]{2015ApJ...799..236N}
{Nita}, G.~M., {Fleishman}, G.~D., {Kuznetsov}, A.~A., {Kontar}, E.~P., \&
  {Gary}, D.~E. 2015, \apj, 799, 236

\bibitem[{{Nita} {et~al.}(2004{\natexlab{a}}){Nita}, {Gary}, \&
  {Lanzerotti}}]{2004SpWea...211005N}
{Nita}, G.~M., {Gary}, D.~E., \& {Lanzerotti}, L.~J. 2004{\natexlab{a}}, Space
  Weather, 2, S11005

\bibitem[{{Nita} {et~al.}(2004{\natexlab{b}}){Nita}, {Gary}, \&
  {Lee}}]{2004ApJ...605..528N}
{Nita}, G.~M., {Gary}, D.~E., \& {Lee}, J. 2004{\natexlab{b}}, \apj, 605, 528

\bibitem[{{Nita} {et~al.}(2017){Nita}, {Viall}, {Klimchuck}, {Loukitcheva},
  {Gary}, {Kuznetsov}, \& {Fleishman}}]{2017ApJ-submitted}
{Nita}, G.~M., {Viall}, N.~M., {Klimchuck}, J.~A., {Loukitcheva}, M.~A.,
  {Gary}, D.~E., {Kuznetsov}, A.~A., \& {Fleishman}, G.~D. 2017, \apj,
  submitted

\bibitem[{{Pesnell} {et~al.}(2012){Pesnell}, {Thompson}, \&
  {Chamberlin}}]{2012SoPh..275....3P}
{Pesnell}, W.~D., {Thompson}, B.~J., \& {Chamberlin}, P.~C. 2012, \solphys,
  275, 3

\bibitem[{{Saint-Hilaire} \& {Benz}(2002)}]{2002SoPh..210..287S}
{Saint-Hilaire}, P., \& {Benz}, A.~O. 2002, \solphys, 210, 287

\bibitem[{{Schou} {et~al.}(2012){Schou}, {Scherrer}, {Bush}, {Wachter},
  {Couvidat}, {Rabello-Soares}, {Bogart}, {Hoeksema}, {Liu}, {Duvall}, {Akin},
  {Allard}, {Miles}, {Rairden}, {Shine}, {Tarbell}, {Title}, {Wolfson},
  {Elmore}, {Norton}, \& {Tomczyk}}]{2012SoPh..275..229S}
{Schou}, J., {et~al.} 2012, \solphys, 275, 229

\bibitem[{{Schwartz} {et~al.}(2002){Schwartz}, {Csillaghy}, {Tolbert},
  {Hurford}, {McTiernan}, \& {Zarro}}]{2002SoPh..210..165S}
{Schwartz}, R.~A., {Csillaghy}, A., {Tolbert}, A.~K., {Hurford}, G.~J.,
  {McTiernan}, J., \& {Zarro}, D. 2002, \solphys, 210, 165

\bibitem[{{Silva} {et~al.}(2000){Silva}, {Wang}, \&
  {Gary}}]{2000ApJ...545.1116S}
{Silva}, A.~V.~R., {Wang}, H., \& {Gary}, D.~E. 2000, \apj, 545, 1116

\bibitem[{{Smith} {et~al.}(2003){Smith}, {Share}, {Murphy}, {Schwartz}, {Shih},
  \& {Lin}}]{2003ApJ...595L..81S}
{Smith}, D.~M., {Share}, G.~H., {Murphy}, R.~J., {Schwartz}, R.~A., {Shih},
  A.~Y., \& {Lin}, R.~P. 2003, \apjl, 595, L81

\bibitem[{{Sui} \& {Holman}(2003)}]{2003ApJ...596L.251S}
{Sui}, L., \& {Holman}, G.~D. 2003, \apjl, 596, L251

\bibitem[{{Takakura}(1972)}]{1972SoPh...26..151T}
{Takakura}, T. 1972, \solphys, 26, 151

\bibitem[{{Wang} {et~al.}(2017){Wang}, {Liu}, {Ahn}, {Xu}, {Jing}, {Deng},
  {Huang}, {Liu}, {Kusano}, {Fleishman}, {Gary}, \&
  {Cao}}]{2017NatAs...1E..85W}
{Wang}, H., {et~al.} 2017, Nature Astronomy, 1, 0085

\bibitem[{{Wheatland} {et~al.}(2000){Wheatland}, {Sturrock}, \&
  {Roumeliotis}}]{2000ApJ...540.1150W}
{Wheatland}, M.~S., {Sturrock}, P.~A., \& {Roumeliotis}, G. 2000, \apj, 540,
  1150

\bibitem[{{White} {et~al.}(2003){White}, {Krucker}, {Shibasaki}, {Yokoyama},
  {Shimojo}, \& {Kundu}}]{2003ApJ...595L.111W}
{White}, S.~M., {Krucker}, S., {Shibasaki}, K., {Yokoyama}, T., {Shimojo}, M.,
  \& {Kundu}, M.~R. 2003, \apjl, 595, L111

\bibitem[{{White} {et~al.}(2011){White}, {Benz}, {Christe},
  {F{\'a}rn{\'{\i}}k}, {Kundu}, {Mann}, {Ning}, {Raulin}, {Silva-V{\'a}lio},
  {Saint-Hilaire}, {Vilmer}, \& {Warmuth}}]{2011SSRv..159..225W}
{White}, S.~M., {et~al.} 2011, \ssr, 159, 225

\bibitem[{{Wiegelmann}(2004)}]{2004SoPh..219...87W}
{Wiegelmann}, T. 2004, \solphys, 219, 87

\bibitem[{{Wiegelmann} {et~al.}(2006){Wiegelmann}, {Inhester}, \&
  {Sakurai}}]{2006SoPh..233..215W}
{Wiegelmann}, T., {Inhester}, B., \& {Sakurai}, T. 2006, \solphys, 233, 215

\end{thebibliography}

\begin{deluxetable}{cc}
\tabletypesize{\scriptsize}
\tablewidth{0pt}
\tablecaption{The summary of the OSPEX fitted parameters for the photon spectrum taken at 18:05:32 UT, with \textit{vth} + \textit{thick2}.}
\tablehead{\colhead{\bf{Parameter}} &  \colhead{\bf{\textit{vth} + \textit{thick2}}}}
\startdata
\cutinhead{\bf{Thermal (\textit{vth})}}
\bf{EM} & 1.3 $\times$ $10^{49}$ cm$^{-5}$ \\
\bf{Temperature} & 2.0 $\times$ $10^7$ K \\
\cutinhead{\bf{Nonthermal}}
 & \bf{thick-target (\textit{thick2})} \\
\bf{{Total integrated electron flux}} & $2.7 \times 10^{35}$ s $^{-1}$ \\
\bf{E$_{cutoff}$} & 22.1 keV \\
\bf{$\delta_{1}$} & 3.3 \\
\bf{E$_{bk}$} & 36 keV\\
\bf{$\delta_{2}$} & 5.5 \\
\bf{E$_{max}$} & 32,000 keV \tablenotemark{\ast} \\
$\chi^2$ & 2.20
\enddata
\tablenotetext{\ast}{The upper energy limit of 32,000 keV was chosen arbitrarily and was fixed during the fit.}
\end{deluxetable}

\begin{landscape}
\begin{deluxetable}{cccccccccccccccccccc}

\tabletypesize{\scriptsize}
\tablecolumns{20}
\tablewidth{0pc}
\tablecaption{Summary of the modeled parameters for all four loops in the final model. The values that have possible ranges are indicated in red, and the ranges are shown in the footnotes.}
\tablehead{
\colhead{Loop} & \colhead{} & \multicolumn{2}{c}{--- B field ---} & \multicolumn{6}{c}{---------------- Thermal Parameters ----------------} & \multicolumn{10}{c}{------------------------------------ Nonthermal Parameters ------------------------------------} \\
\colhead{Designation} & \colhead{Radius} & \colhead{B$_{max}$} & \colhead{B$_{min}$} & \colhead{$n_{0}$} & \colhead{T} & \colhead{$p_{0}$} & \colhead{$p_{1}$} & \colhead{$q_{0}$} & \colhead{$q_{2}$} & \colhead{$n_{b}$} & \colhead{$p_{0}$} & \colhead{$p_{1}$} & \colhead{$q_{0}$} & \colhead{$q_{2}$} & \colhead{E$_{cutoff}$} & \colhead{$\delta_{1}$} & \colhead{E$_{break}$} & \colhead{$\delta_{2}$} & \colhead{E$_{max}$} \\
\colhead{} & \colhead{(grid pts)} & \colhead{(G)} & \colhead{(G)} & \colhead{(cm$^{-3}$)} & \colhead{(MK)} & \colhead{} &\colhead{} & \colhead{} & \colhead{} & \colhead{(cm$^{-3}$)} & \colhead{} & \colhead{} & \colhead{} & \colhead{} &  \colhead{(keV)} & \colhead{} & \colhead{(keV)} & \colhead{} & \colhead{(MeV)}}
\startdata
\bf{Thermal-only} & 5 & 856 & 1925 & $1.1-1.6 \times 10^{11}$ & 20 & 0.6 & 3 & 1.3 \tablenotemark{\ast} & -0.28 \tablenotemark{\ast} & 0 & - & - & - & - & - & - & - & - & - \\
\bf{Lower} & 10 & 401 & 1847 & \textcolor{red}{$5.0 \times 10^{9}$} \tablenotemark{\bigstar} & \textcolor{red}{20} \tablenotemark{\bigstar} & 2 & 2 & - & - & $3.0 \times 10^{7}$ & 2 & 2 & 3.9 & 0.02 & 22.1 & 5.0 & \textcolor{red}{200} \tablenotemark{\spadesuit} & 3.0 & 10 \\
\bf{Higher} & 10 & 250 & 1984 & \textcolor{red}{$5.0 \times 10^{9}$} \tablenotemark{\bigstar} & \textcolor{red}{20} \tablenotemark{\bigstar} & 2 & 2 & - & - & \textcolor{red}{$4.5 \times 10^{8}$} \tablenotemark{\clubsuit} & 2 & 2 & 15 & 0 & 22.1 & 3.6 & 43 & 6.8 & 10 \\
\bf{Overarching} & 23 & 85 & 225 & \textcolor{red}{$5.0 \times 10^{9}$} \tablenotemark{\bigstar} & \textcolor{red}{20} \tablenotemark{\bigstar} & 2 & 2 & - & - & \textcolor{red}{$6.1 \times 10^{6}$} \tablenotemark{\diamondsuit} & 3.5 & 3.5 & 16 & 0 & \textcolor{red}{22.1} \tablenotemark{\diamondsuit} & 2.5 & - & - & 5
\enddata

\tablenotetext{\ast}{Modified from the default function by adding a simple Gaussian-like function used for nonthermal population (See Appendix Eqn. B4).}

\tablenotetext{\bigstar}{\textbf{The lower and the higher loop: $\mathbf{T < 50}$ MK and $\mathbf{n_{0} < 2 \times 10^{10}}$ cm $\mathbf{^{-3}}$. The overarching loop: $\mathbf{T < 30}$ MK and $\mathbf{n_{0} < 2 \times 10^{10}}$ cm$\mathbf{^{-3}}$.} The thermal parameters for these loops are not strictly constrained since the emission measure is solely constrained by the thermal-only loop. See section 4.3 for details.}

\tablenotetext{\clubsuit}{\textbf{Range: $\mathbf{\sim 10^{8} < n_{b} < 2 \times 10^{10}}$ cm$\mathbf{^{-3}}$.} The lower limit is considered based on the upper limit of the thermal density, while the upper limit is equal to the upper limit of the thermal density. See section 4.3 for details.}

\tablenotetext{\diamondsuit}{\textbf{Range: $\mathbf{4 \times 10^{4} < n_{b} < 6.7 \times 10^{6}}$ cm$^{\mathbf{-3}}$, with $\mathbf{E_{cutoff} \sim 600 keV}$ corresponding to $\mathbf{n_{b} \sim 4 \times 10^{4}}$.} The nonthermal electron energy spectrum for the overarching loop cannot be constrained under several hundred keVs since the emission from this loop is only visible in the MW low-frequency range. See Section 4.3 for details.}

\tablenotetext{\spadesuit}{\textbf{Range: $\mathbf{180 keV \lesssim E_{break} \lesssim 220 keV}$.} The choice of E$_{break} = 200$ keV is arbitrary, since this is above the statistically significant energy range of the HXR observation (67 keV). See Section 4.3 for details.}

\end{deluxetable}
\end{landscape}

\end{document}